\documentclass[prd,preprintnumbers,preprint,amsmath,amssymb,nofootinbib,nopacs]{revtex4}
\usepackage{latexsym,ifthen,graphics,color,epsfig,hyperref}

\newcommand{\ie}{{i.e.}}

\newcommand{\eg}{{e.g.}}

\newcommand{\viz}{{viz.}}
\newcommand{\wrt}{with respect to}
\newcommand{\lhs}{left-hand side}

\newcommand{\rhs}{right-hand side}

\newcommand{\be}{\begin{equation}}
\newcommand{\ee}{\end{equation}}
\newcommand{\bea}{\begin{eqnarray}}
\newcommand{\eea}{\end{eqnarray}}
\newcommand{\beas}{\begin{eqnarray*}}
\newcommand{\eeas}{\end{eqnarray*}}

\newcommand{\bear}{\begin{array}{l}}
\newcommand{\eear}{\end{array}}

\newcommand{\bcf}{\begin{center}\begin{figure}}
\newcommand{\ecf}{\end{figure}\end{center}}

\newcommand{\bct}{\begin{center}\begin{table}}
\newcommand{\ect}{\end{table}\end{center}}

\newcommand{\ds}{\displaystyle}

\newcommand{\eq}[1]{(\ref{eq:#1})}
\newcommand{\eqn}[1]{equation~(\ref{eq:#1})}

\newcommand{\eqs}[2]{(\ref{eq:#1}) and~(\ref{eq:#2})}

\newcommand{\sect}[1]{section~\ref{sec:#1}}
\newcommand{\Sect}[1]{Section~\ref{sec:#1}}

\newcommand{\app}[1]{appendix~\ref{app:#1}}
\newcommand{\App}[1]{Appendix~\ref{app:#1}}

\newcommand{\Int}[1]{\int \!\! d^D \! #1 \,}
\newcommand{\FourInt}[1]{\int \!\! d^4 \! #1 \,}
\newcommand{\TwoInt}[1]{\int \!\! d^2 \! #1 \,}

\newcommand{\MomInt}[2]{\int \!\! \frac{d^{#1} #2}{(2\pi)^{#1}} \, }

\newcommand{\der}[2]{\frac{d #1}{d #2}}
\newcommand{\pder}[2]{\frac{\partial #1}{\partial #2}}

\newcommand{\fder}[2]{\ensuremath{\frac{\delta #1}{\delta #2}}}

\newcommand{\hf}{\frac{1}{2}}

\newcommand{\measure}[1]{\mathcal{D} #1 \, }

\newcommand{\one}{1\!\!1}

\newcommand{\Tr}{\mathrm{Tr}\,}

\newcommand{\nCr}[2]{
		\Bigl(
		\begin{matrix}
			#1
		\\[-2ex]
			#2
		\end{matrix}
		\Bigr)
}

\newcommand{\norm}[1]{
	\lVert #1 \rVert	
}

\newcommand{\inner}[2]{
	\langle #1,#2 \rangle
}

\newcommand{\dd}{\dot{\Delta}}
\newcommand{\td}{\tilde{\Delta}}
\newcommand{\knl}[1]{\cdot {#1}\cdot}

\newcommand{\critexp}{\lambda}

\newcommand{\flow}{\Lambda \partial_\Lambda}
\newcommand{\thetaflow}{\thetabar \partial_{\thetabar}}

\newcommand{\Sint}{S^{\mathrm{int}}}

\newcommand{\hS}{\hat{S}}

\newcommand{\dual}{\mathcal{D}}
\newcommand{\dualv}[1]{\dual^{(#1)}}

\newcommand{\thetabar}{\overline{\theta}}
\newcommand{\nubar}{\overline{\nu}}
\newcommand{\chibar}{\overline{\chi}}

\newcommand{\numX}{\xi}

\newcommand{\classical}[3]{\pder{#1}{\phi} \knl{#2} \pder{#3}{\phi}} 
\newcommand{\quantum}[2]{\pder{}{\phi} \knl{#1} \pder{#2}{\phi}}

\newcommand{\expoplam}{
	\exp
	\left(
		- \frac{\hat{\lambda}}{2\chi_4} \pder{}{\phi} \knl{\Delta} \pder{}{\phi}
	\right)
}

\newcommand{\abar}{\bar{a}}
\newcommand{\SGW}{S_{\mathrm{GW}}}

\newcommand{\arXiv}[2]{arXiv:{#1} [#2]}

\newcommand{\hepth}[1]{hep-th/#1}

\begin{document}

\title{Wilsonian Renormalization of Noncommutative Scalar Field Theory}

\author{Razvan Gurau}
\affiliation{Perimeter Institute for Theoretical Physics, Waterloo, ON, N2L 2Y5, Canada}
\email{rgurau@perimeterinstitute.ca}
\author{Oliver~J.~Rosten}
\affiliation{Department of Physics and Astronomy, University of Sussex, Brighton, BN1 9QH, U.K.}
\email{O.J.Rosten@Sussex.ac.uk}

\begin{abstract}

Drawing on analogies with the commutative case, the Wilsonian picture
of renormalization is developed for noncommutative scalar field theory.
The dimensionful noncommutativity parameter, $\theta$, induces several new features. Fixed-points are replaced by `floating-points' (actions which are scale independent only up to appearances of $\theta$
written in cutoff units). Furthermore, it is found that one must use correctly normalized operators, with respect to a new scalar product, to
define the right notion of relevance and irrelevance.

In this framework it is straightforward and intuitive to reproduce the classification of operators found by Grosse \& Wulkenhaar, around the Gaussian floating-point. The one-loop $\beta$-function of their model
is computed directly within the exact renormalization group, reproducing the previous result that it vanishes in the self-dual theory, in the limit of large cutoff. With the link between this methodology and earlier results made, it is discussed how the vanishing of the $\beta$-function to all loops, as found by Disertori et al., should be  interpreted in a Wilsonian framework.

\end{abstract}

\maketitle

\tableofcontents

\section{Introduction}

\subsection{The Grosse \& Wulkenhaar Model}

The renormalization of noncommutative theories---which not only appear naturally in some limiting regime of string theory~\cite{Connes:1997cr} but are also relevant for physics in a strong magnetic 
field~\cite{Douglas:2001ba}---is a subtle and instructive business. 

Whilst one might
hope that a fundamental graininess of spacetime would alleviate the ultraviolet divergences ubiquitous in quantum field theory, instead one encounters the infamous ultraviolet/infrared (UV/IR) mixing 
problem~\cite{NCPD}.  Compared with commutative theories, their noncommutative analogues exhibit a new class of  graph, like the non-planar tadpole in a $\phi^4$ theory.
Whilst UV convergent, these graphs blow up for vanishing external momenta, if the overall UV cutoff is removed. Embedding these non-planar diagrams inside bigger graphs can 
ultimately generate loop integrals which are not integrable, due to \emph{IR} divergences.

Thus, although this UV/IR mixing appears at the first order in perturbation theory, graphs with divergent amplitude due to this effect appear only at higher orders (at least three for a four dimensional theory). This initially led to some confusion about the renormalizability of such theories~\cite{C+R-1,C+R-2}. We know today that na\"{\i}ve noncommutative theories in fact are not renormalizable but, in order to see this, one needs to go beyond the first orders of perturbation theory, and address the question of renormalizability to all orders or nonperturbatively.\footnote{Throughout this paper, nonperturbative is taken in the Wilsonian, or `exact renormalization group' sense. This is by no means equivalent to the meaning of nonperturbative in the constructive quantum field theory sense. Thus we do not claim to prove, for instance, that the perturbative series is Borel summable.}

Remarkably, despite the apparently hopeless situation caused by UV/IR mixing,  Grosse \& 
Wulkenhaar were able to show, in a trail-blazing series of papers~\cite{G+W-2D,G+W-PC,G+W-4D}, that a particular version of the scalar $\phi^4$ model on noncommutative $\mathbb{R}^4$ \emph{is} renormalizable. 
The initial, highly technical proofs---which were performed in the `matrix base', that will be introduced shortly---have been simplified and translated to the direct space in \cite{Gurau:2005gd}.
The key step in Grosse \& Wulkenhaar's approach was to
add a new two-point term to the action. Heuristically, the motivation for this can be justified in several ways.  Physically, Grosse \& Wulkenhaar recognized that the message of UV/IR mixing is precisely that the short distance physics of the model modifies the long distance physics. In the commutative $\phi^4$ model, the long distance physics corresponds to an essentially free theory. Thus, to have any hope of renormalizing the noncommutative model, they argued that the free theory should be modified; this of course amounts to adding a new two-point term to the action. The second justification comes from a duality of certain noncommutative interactions first recognized by Langmann \& Szabo~\cite{L-S-duality}. 

To precisely state the Grosse \& Wulkenhaar modifications we introduce the following notation.
The coordinates of the noncommutative Moyal space $\mathbb{R}^4_\theta$ satisfy the commutation relation
\be
	[x^\mu, x^\nu] = i\theta^{\mu\nu},
\ee
where we will choose a  coordinate system in which the antisymmetric matrix $\theta^{\mu\nu}$ takes the form
\be
	\theta_{\mu \nu} =
	\left(
		\begin{array}{cccc}
			0 & \theta_1 & 0 & 0
		\\
			-\theta_1 & 0 & 0 & 0
		\\
			0 & 0 & 0 & \theta_2
		\\
			0 & 0 & -\theta_2 & 0
		\end{array}
	\right).
\ee
For the rest of this paper we will only consider the case that
\be
	\theta_1 = \theta_2 = \theta;
\ee
it is straightforward to adapt our methodology and conclusions to the general case.

The algebra of functions on $\mathbb{R}^4_\theta$ can be taken as the algebra of Schwarz-class functions of rapid decay on the usual $\mathbb{R}^4$, but with a deformed product:
\be
	(a\star b)(x) = \MomInt{D}{k} \Int{y} 
	a(x+{\scriptstyle \hf} \theta \cdot k)
	b(x+y) e^{ik \cdot y},
\label{eq:star}
\ee
where
\[
	\theta \cdot k \equiv \theta^{\mu\nu} k_\nu, 
	\qquad k\cdot y \equiv k^\mu y_\mu, 
	\qquad
	\theta^{\mu\nu} = -\theta^{\nu\mu}.
\]

The Grosse \& Wulkenhaar action is
\be
	\SGW = \FourInt{x}
	\left[
		\hf 
		\left(
			\partial_\mu \phi
		\right)
		\star
		\left(
			\partial^\mu \phi
		\right)
		+\frac{\Omega^2}{2}
		\left(
			\tilde{x}_\mu \phi
		\right)
		\star
		\left(
			\tilde{x}^\mu \phi
		\right)
		+
		\frac{M^2}{2} \phi \star \phi
		+\frac{\lambda}{4!} \phi \star \phi \star \phi \star \phi
	\right],
\label{eq:G+W}
\ee
where
\be
	\tilde{x}_\mu \equiv 2 (\theta^{-1})_{\mu\nu} x^{\nu}.
\ee
The new two-point interaction takes the form of a harmonic oscillator term. At first sight, this is a
rather pathological addition to the action, since it breaks translation invariance. However, we are
viewing the Grosse \& Wulkenhaar model simply as a laboratory for understanding renormalization in noncommutative field theories. As pointed out by Rivasseau~\cite{Riv-Review} there are other models
in which, whilst a similar sort of term must be added, physical observables turn out not to feel the translation invariance violating effects. So we hope that the results of this paper will be adaptable to more physically interesting cases.

With this issue behind us, we now return to Langmann--Szabo duality. They observed that an interaction of the form $\phi\star \cdots \star \phi$ is invariant under an interchange of positions and momenta. In other words, the 
Fourier transformed interaction takes the same form as the position-space interaction. Specifically, such $\star$-product interactions are invariant under
\be
	\hat{\phi}(p) \leftrightarrow \pi^2 \sqrt{\lvert \det \theta \rvert} \phi(x),
	\qquad 
	p_\mu \leftrightarrow \tilde{x}_\mu,
\ee
where $\hat{\phi}(p_a) = \FourInt{x} e^{(-1)^a i p_{a,\mu} x^\mu_a} \phi(x_a)$, with the label, $a$, indicating the position of the $\phi$ within the string  $\phi\star \cdots \star \phi$.

Now, returning to $\SGW$, both the mass term and the four-point term are invariant under the duality transformation, but the kinetic term is not. Adding the harmonic oscillator term implements this duality for the complete action, at least for $\Omega=1$, the so-called `self-dual' theory. Away from the self-dual point, the theory is covariant (rather than invariant) under the Langmann-Szabo duality:
\be
	S[\phi](\Omega,M,\lambda) \mapsto 
	\Omega^2 \, S[\phi]\Bigl(\frac{1}{\Omega},\frac{M}{\Omega},\frac{\lambda}{\Omega^2}\Bigr).
\ee
Grosse \& Wulkenhaar argued that if the standard kinetic term has divergent contributions, then it is likely that its dual will also pick up divergent contributions. If these are to be absorbed by tuning the bare action, this demands that the harmonic oscillator term be present from the start.
So, one way or another, the addition of the harmonic oscillator term is plausible if we are to construct a renormalizable theory.

To actually demonstrate perturbative renormalization of their theory, Grosse \& Wulkenhaar adapted Polchinski's proof~\cite{Pol} of the perturbative renormalizability of the commutative $\phi^4$ model. This proof uses an exact renormalization group (ERG) equation (often called a flow equation), in the spirit of Wilson~\cite{Wilson} (see also~\cite{WH}). The basic idea of the formalism is that, starting at some high energy scale---the bare scale---$\Lambda_0$, one integrates out degrees of freedom down to a much lower `effective' scale $\Lambda$. During this procedure, the bare action evolves into the Wilsonian effective action, $S_\Lambda$. Flow equations provide an exact equation specifying how $S_\Lambda$ changes with scale, \viz
\[
	\partial_t S_\Lambda = \ldots,
\]
where $t$ is the `RG-time' which, given an arbitrary (as opposed to physical) scale, $\mu$, is defined according to $t \equiv \ln \mu /\Lambda$.
One of the particularly useful aspects of Polchinski's work is a flow equation (see \sect{Flow}) that is considerably simpler to use than Wilson's.

Nevertheless, at first sight it is quite remarkable that a flow equation approach works in noncommutative scalar field theory. A crucial ingredient of flow equations is that coarse-graining of degrees of freedom occurs only over a local patch. This can be achieved by ensuring that all ingredients of the flow equation have a derivative expansion, which is necessary to ensure that each RG step $\Lambda \mapsto \Lambda - \delta \Lambda$ is free of IR divergences. In the noncommutative case, it looks like this approach is doomed to failure. Fortunately, however, it turns out that field theories on $\mathbb{R}^4_\theta$ can be reformulated in terms of infinite dimensional matrices~\cite{MatrixBase} (see~\cite{G+W-2D} for a digestible summary). In the matrix base, there is no problem writing down a flow equation and, indeed, this is precisely the approach taken by Grosse \& Wulkenhaar.\footnote{Presumably, once things are formulated in the matrix base, it is possible (but beyond the scope of this paper) to translate back to the direct space.}

To construct the matrix base (which is reviewed in more detail in \app{basis}) in $D=4$, we start with the basis function $b_{mn}(x)$. This is built from two copies of the $D=2$ basis functions and, as a consequence, $m$ and $n$ are valued in $\mathbb{N}^2$:
\be
	b_{mn}(x) = f_{m^1n^1}(x^1,x^2)f_{m^2n^2}(x^3,x^4),
	\qquad
	m = \begin{matrix} m^1 \\[-1ex]  m^2 \end{matrix} \in \mathbb{N}^2,
	\quad
	n = \begin{matrix} n^1 \\[-1ex]  n^2 \end{matrix} \in \mathbb{N}^2,
\label{eq:adapt}
\ee
where the $m^i$ and $n^i$ are just valued in $\mathbb{N}$.

We can now express functions in the direct space in the matrix base:
\be
	\phi(x) = \sum_{m,n=0}^\infty \phi_{mn} b_{mn}(x).
\ee

One of the nice things about the matrix base is that the $\star$-product translates to matrix multiplication:
\be
	(\phi\star\phi)(x) = \sum_{m,n,k=0}^\infty \phi_{mk} \phi_{kn} b_{mn}(x).
\ee
Since
\be
	\FourInt{x} b_{mn}(x) = \nu_4 \delta_{mn},
\ee
where $\nu_4$---the volume of an elementary cell---is defined according to
\be
	\nu_4 \equiv (2\pi\theta)^2,
\ee
the four-point interaction is particularly simple in the matrix base:
\be
	\FourInt{x} (\phi\star\phi\star\phi\star\phi)(x) = 
	\nu_4 \sum_{m,n,k,l} \phi_{mn} \phi_{nk} \phi_{kl} \phi_{lm}.
\ee

To deal with the kinetic term, we recognize that both derivatives \wrt\ $x_\mu$ and pointwise multiplication by $\tilde{x}_\mu$ can be written in terms of $\star$-products~\cite{Szabo-Rev}
(see \app{basis}):
\be
	\tilde{x}_\mu f(x) = \hf \{ \tilde{x}_\mu , f\}_\star(x),
	\qquad
	\pder{f}{x_\mu} = -\frac{i}{2} [\tilde{x}_\mu, f]_\star (x).
\label{eq:replace}
\ee
Thus, defining the matrix version of $\tilde{x}_\mu$ via
\be
	\tilde{x}_\mu = \sum_{m,n=0}^\infty \bigl( \tilde{X}_{\mu} \bigr)_{mn} b_{mn}(x)
\ee
and splitting $\Omega^2$ up into the self-dual part and the deviation from this:
\be
	\Omega^2 = 1+\omega,
\ee
we can write $\SGW$ as:
\be
	\SGW = \nu_4
	\left[
		\frac{1+\omega/2}{2} \phi \cdot \tilde{X}_\mu \cdot \tilde{X}^\mu \cdot \phi
		+\frac{\omega}{4} 
		\phi \cdot \tilde{X}_\mu  \cdot \phi \cdot \tilde{X}^\mu
		+
		\frac{M^2}{2} \phi \cdot \phi +\frac{\lambda}{4!} \phi  \cdot \phi  \cdot \phi  \cdot \phi
	\right],
\ee
where the dot-notation is shorthand for matrix summations:
\[
	\phi  \knl{Y} \phi
	\equiv
	\sum_{m,n,k,l \epsilon \mathbb{N}^2}
	\phi_{mn} Y_{mn;kl} \phi_{kl}.
\]
Note that we are using the same symbol for the field in position space and in the matrix base, but it should always be clear from the context which is which.

For all that follows, it will be very convenient to consider the first term to be the kinetic term, with the other two-point terms thought of as perturbations. The primary advantage of this is that this kinetic term is readily invertible; the (bare) propagator defined as the inverse of the full two-point function is rather unpleasant~\cite{G+W-4D}. So, writing
\be
	 \phi \cdot \tilde{X}_\mu \cdot \tilde{X}^\mu \cdot \phi = \sum_{m,n,k,l} G_{mn;kl} \phi_{mn} \phi_{kl},
\ee
it turns out that
\be
	G_{mn;kl} = \frac{4(2+m+n)}{\theta} \delta_{ml}\delta_{nk}.
\label{eq:CTP}
\ee

We conclude this section by giving an expression for the matrices $\tilde{X}_\mu$. Actually, henceforth, we will define $\tilde{X}_\mu$ such that it has been rendered dimensionless by the appropriate power of the effective scale, $\Lambda$:
\be
	\tilde{X}_1 = \tilde{X}_3 =
	\sqrt{\frac{2}{\thetabar}}
	\begin{pmatrix}
		0 & 1 & &
	\\
		1 & 0 & \sqrt{2} &
	\\
		 & \sqrt{2} & 0 & \ddots
	\\
		&& \ddots & \ddots
	\end{pmatrix},
	\qquad
	\tilde{X}_2 = \tilde{X}_4 =
	i
	\sqrt{\frac{2}{\thetabar}}
	\begin{pmatrix}
		0 & -1 & &
	\\
		1 & 0 & -\sqrt{2} &
	\\
		 & \sqrt{2} & 0 & \ddots
	\\
		&& \ddots& \ddots
	\end{pmatrix},
\label{eq:tildeX}
\ee
where the above matrices are tensor products with the identity $\delta_{m^2n^2}$ for $\tilde{X}_1$ and $\tilde{X}_2$ and with $\delta_{m^1n^1}$ for $\tilde{X}_3$ and $\tilde{X}_4$. 
The only thing we need for this paper is to notice that these expressions contain a $1/\sqrt{\thetabar}$, with $\thetabar$ defined in eq. (\ref{eq:thetabar}).

\subsection{Renormalizability}
\label{sec:renorm}

As mentioned already, the defining achievement of Grosse \& Wulkenhaar was to prove perturbative renormalizability of their model. That this was achieved using flow equations is tantalizing since, in the
commutative case, flow equations provide a natural framework for concretely accessing Wilson's picture~\cite{Wilson,TRM-Elements} of nonperturbative renormalizability.

In the commutative setting, rather than seeking to prove perturbative renormalizability, one can start by classifying nonperturbatively renormalizable theories by looking for critical fixed-points. The rationale for doing this (which will be discussed in more detail in \sect{WR}) is that such theories are conformal. Conformal theories are independent of all scales and, therefore, are independent of the bare scale, $\Lambda_0$. Consequently, $\Lambda_0$ can be safely (indeed, trivially) sent to infinity: the theory is nonperturbatively renormalizable.
Having found a conformal field theory, scale dependent renormalizable theories can be found by perturbing the fixed-point action in the directions relevant (including marginally relevant) to the given fixed-point (again, this will be discussed in more detail in \sect{WR}).
However, the main point that we wish to elucidate in this section is that this 
necessitates looking at quantum field theories (QFT)s in a somewhat different way from usual. 

The typical starting place in QFT is some action which,  for whatever reason, we have decided to investigate.  However, if we are interested in classifying nonperturbatively renormalizable theories, our starting point is not a specific action, but rather the `theory space' consisting of \emph{all possible} actions (limited by certain requirements to be discussed in \sect{Eff-Ac}).  As stated above, we are interested (in the first instance) in picking out those actions corresponding to critical fixed-points.
Thus, in this way of looking at things, the actions with which we work are not given to us ahead of time: they are objects for which we \emph{solve}. In the case of asymptotically free theories, this procedure is essentially trivial since we are dealing either with the Gaussian theory and the weakly coupled (renormalizable) theories in its vicinity. In such cases, one can dispense with the Wilsonian approach and jump straight to the actions which are renormalizable by power counting. Nevertheless, it should be emphasised that by glossing over the underlying Wilsonian picture, details such as why the perturbative renormalizability of $\lambda \phi^4$ theory in $D=4$ by no means implies nonperturbative renormalizability are obscured.
(The answer resides in the fact that, since $\lambda$ is marginally \emph{irrelevant}, it is not a relevant perturbation of the Gaussian fixed-point: if a  theory well approximated by a $\lambda \phi^4$ action at low energies is to be nonperturbatively renormalizable, it must be sitting on an RG trajectory which emanates, in the UV, from some non-trivial fixed-point.%
\footnote{Alternatively, in the constructive approach this is seen as follows. When sending the UV cutoff to $\infty$ the bare coupling constant flows out of the circle of Borel summability of the perturbative series.}
As it happens, such a scenario was ruled out in~\cite{Trivial}).

In this paper, our aim is to apply this Wilsonian way of thinking to scalar field theory on 
$\mathbb{R}^4_\theta$. Again, we emphasise that our aim is not to apply the Wilsonian way of thinking
to the Grosse \& Wulkenhaar model, per se. This would amount to putting the coach before the horse! Rather, we want our flow equation to tell us which theories we should be looking at, if we are interested in things which are renormalizable \emph{beyond} perturbation theory. In the process of doing this,
we will discover a number of novelties. The first thing we find is that fixed-points are no longer of any use to us! This is because independence of an action\footnote{We assume that we are working in dimensionless variables---see \sect{WR}.} on $\Lambda$ no longer implies a theory which is necessarily independent of the bare scale. This is due to the presence of the dimensionful quantity $\theta$, from which we can construct the dimensionless, $\Lambda$ independent object
\[
	\theta \Lambda_0^2.
\]
However, as we will discuss further in \sect{Float}, this problem can be removed if, instead of considering fixed-points, we instead look for \emph{floating-points}: points which are independent of the scale, up to dependence on
\be
	\thetabar \equiv \theta \Lambda^2.
\label{eq:thetabar}
\ee

Having found a floating-point, the next thing to do---just as in the commutative case---is to linearize the ERG equation about the floating point and find the eigenoperators, $\mathcal{O}$, and the associated RG eigenvalues. The RG eigenvalue of an operator determines whether it is relevant or irrelevant (an operator may be marginal, in which case one must go to higher order to determine whether it is marginally relevant or marginally irrelevant). Here we encounter the second novelty, again due to the presence of $\theta$. In the vicinity of a floating-point, with action $S_*$, we write the action as:%
\footnote{The star, $*$, will be used to denote either floating-point or fixed-point quantities and has nothing to do with the $\star$-product.}
\be
	S[\phi](t,\thetabar) = S_*[\phi](\thetabar) + \sum_i g_i(t,\thetabar) \mathcal{O}_i[\phi],
\ee
where the $g_i$ are the couplings. 

Now, it turns out that there exists a natural inner product on the space spanned by the $\mathcal{O}_i$ (we will define what we mean by natural in a moment). 
With this in mind, we perform the trivial rewriting:
\be
	S_t = S_* + \sum_i g_i(t,\thetabar) \norm{\mathcal{O}_i} \hat{\mathcal{O}}_i.
\ee
The key point is that $ \norm{\mathcal{O}_i}$ depends on $\thetabar$. Thus, defining
\be
	\hat{g}_i \equiv g_i \norm{\mathcal{O}_i}
\ee
it is clear that, in general, the $\hat{g}_i$ will have different dependence on $\Lambda$ than the
$g_i$. This is crucial when determining (ir)relevance: only when we use the couplings conjugate to the
$\hat{\mathcal{O}}_i$ do we obtain the correct results. (We emphasise that there is no analogue of this in the commutative case, where any norm of the eigenoperators would be independent of scale, and so uninteresting.)

At first sight, however, the inner product that we take is just one possible choice and other choices will
give different normalizations and hence a different classification of (ir)relevance. But there is a constraint that we should place on the inner product to do with reparametrization invariance~\cite{WegnerInv,TRM-Elements}. By analogy with the commutative case we expect (and, indeed, will explicitly show in the Gaussian case in \sect{Float}) that every floating-point belongs to a line of \emph{equivalent} floating-points, related to each other by a reparametrization of the field. A manifestation of this is the presence of an exactly marginal (redundant) direction which takes us from one representative of the floating-point to the rest. 

We now demand that our inner product is such that the operator which maps us between equivalent representations of our theory is exactly marginal, at all scales. However, given the spectrum of operators at the Gaussian floating-point (which we calculate in \sect{NCRTs}) it is only at the Gaussian floating-point that
we know precisely what this operator is, without further calculation. It turns out  that this operator does, indeed, come out marginal with our initial guess at an inner product, but only in the large $\Lambda$-limit. 
That it works in this limit is rather fortunate, since we are ultimately interesting in trajectories which sink back into the Gaussian floating-point precisely as $\Lambda \rightarrow \infty$ (these being the noncommutative version of Wilson's renormalized trajectories~\cite{Wilson}).

Consequently, we have at our disposal an inner product which is useful only for classifying perturbations  of the Gaussian floating-point spawned in the large-$\Lambda$ limit.
With this in mind, we assess (ir)relevance by looking at the behaviour of
\be
	\lim_{\Lambda \rightarrow \infty} \hat{g}_i(t,\thetabar) 
	=  \lim_{\Lambda \rightarrow \infty} g_i(t,\thetabar)  \norm{\mathcal{O}_i}.
\ee

In this way of looking at things, we are able to very easily recover the conclusion that, at linear order, the relevant and marginal operators of the Gaussian floating-point are as follows: the mass is relevant and both $\hat{\omega}$ and $\hat{\lambda}$ are marginal. That this comes about easily is in fact an achievement. 
For example, the mass vertex can be written
\[
	\frac{M^2}{2} \sum_{m,n,k,l}
	\phi_{mn} \phi_{kl} A_{mn;kl},\qquad	\mbox{with}\ A_{mn;kl} = \delta_{ml} \delta_{nk}.
\]
But if this operator is relevant (as we would expect, by power counting), then why aren't operators with more complicated versions of $A_{mn;kl}$ also relevant? In our picture, this is because these more exotic terms have $\thetabar$ dependent norms, which alters what we should take as the coupling. 
A similar conclusion as to which operators are relevant/marginal was obtained by Grosse \& Wulkenhaar in~\cite{G+W-4D}, but we believe that our approach is much more transparent.

As in the commutative case, to determine whether the marginal operators are really relevant or irrelevant, we must go beyond leading order. Equivalently, for the Gaussian floating-point, where the interesting question is the (ir)relevance of the four-point coupling, $\hat{\lambda}$, we can just perform a standard calculation of the $\beta$-function. This is done in \sect{beta}, and we recover Grosse \& Wulkenhaar's result that,
\emph{for the self-dual theory},
\be
	\lim_{\mathcal{N}\rightarrow\infty} \beta \equiv \lim_{\mathcal{N}\rightarrow\infty}
	\mathcal{N} \der{\lambda}{\mathcal{N}} = 0, \ \mbox{at one loop}.
\label{eq:one-loop-vanish}
\ee
where $\mathcal{N} \sim \thetabar $ is a cutoff on the matrix indices (note the $\lambda$ and $\hat{\lambda}$ are the same in the large-$\Lambda$ limit).
It is nice to see how their calculation can be done directly (and simply) within the ERG and, indeed, it is obvious that our calculation can be mapped onto the earlier one.

The one loop result has been extended by Disertori et al.~\cite{Vanishing} to all orders in perturbation theory:
\bea
	\mathcal{N} \frac{d\lambda}{d\mathcal{N}}= O\Bigl(\frac{\ln \mathcal{N}}{\mathcal{N}^2}\Bigr)\; ,
\eea
One then concludes that, perturbatively,
\bea
\lambda(\mathcal{N})-\lambda^{UV}\approx\int_{\mathcal{N}}^{\infty}d\mathcal{N}' \frac{\ln \mathcal{N}'}{(\mathcal{N}'+2)^3} <\frac{1}{\mathcal{N}+2} \; .
\eea
Thus, perturbatively (to all orders), the flow of the coupling constant is finite. This, in turn, is the  traditional perturbative signature of asymptotic safety. 
Of course, it is quite possible that  the exact marginality of $\hat{\lambda}$ can be violated by exponentially small terms, such as $e^{-1/\hat{\lambda}}$, which are perturbatively (even at all orders) invisible. In this case, the determination of whether or not the Gaussian floating-point supports asymptotically free trajectories is, interestingly, a nonperturbative one. (We will discuss in a moment
possible consequences of the difference between $\lambda$ and $\hat{\lambda}$.)

Now let us consider how to interpret the result of Disertori et al.\ if it happens to turn out 
 that $\hat{\lambda}$ is exactly marginal.
 In a \emph{commutative} theory, an exactly marginal (non-redundant) direction would suggest a line of (inequivalent) fixed-points, connected to the Gaussian one.%
\footnote{
Note that, within the ERG, one would have to \emph{compute} the form of these fixed-points since a
fixed-point action, plus a marginal perturbation is only a fixed-point action at linear order in
the perturbation. In the absence of a better alternative, one could compute the non-trivial fixed-point
action using a perturbation series in the exactly marginal coupling.
}
 In other words, in the scenario that ${\lambda}$ were marginal (still in the commutative setting), we would have a line of conformal field theories labelled by the strength of ${\lambda}$ (of course, this is not realized in practise, since ${\lambda}$ is already found not to be marginal at one-loop). 

In the noncommutative case, however, marginality of $\hat{\lambda}$ does not necessarily imply a line of inequivalent \emph{floating}-points. 
The reason for this is that a marginal coupling is one for which
\be
	\lim_{\Lambda \rightarrow \infty} \Lambda \der{\hat{g}_i(t,\thetabar)}{\Lambda} = 0,
\label{eq:marginal-criterion}
\ee
whereas our criterion that a deformation of a floating-point is another floating point is
\be
	\Lambda \left.\pder{\hat{g}_i(t,\thetabar)}{\Lambda}\right\vert_{\thetabar} = 0.
\label{eq:floating-criterion}
\ee

Before going any further, we should note that it is not currently clear whether the presence of the limit 
in~\eq{marginal-criterion} is an artefact of our limited understanding of how to define the couplings in the noncommutative theory, or a real effect. To recap: to properly define the couplings requires that we have an appropriate norm and, at the moment, we only know how to construct this in the large-$\Lambda$ limit, in the vicinity of the Gaussian floating-point. If we had a norm with a larger range of validity then we could, in principle, compute the $\beta$-function at all scales. Of course, this would not necessarily mean that any change would need to be made to~\eq{marginal-criterion}---we might find that the $\beta$-function does, indeed, only vanish in the large-$\Lambda$ limit.

As we will see in \sect{NCRTs}, the coupling conjugate to the operator $\mathcal{O}_j$ takes the form,
to linear order in perturbations about the  Gaussian floating-point,
\[
	\hat{g}_j[\alpha](t, \thetabar) \sim \int \! ds \, \alpha_j(s) e^{\zeta_j t} \thetabar^s
	\norm{\mathcal{O}_j},
\]
where $\zeta_j - 2s$ is an integer (the analogue of the RG eigenvalue) and $\alpha_j(s)$ is an integration constant. For the coupling $\hat{\lambda}$, 
we will show that
\emph{in the limit $\Lambda \rightarrow \infty$} (i)
the norm is independent of scale (ii) at linear order in the perturbation away from the Gaussian floating-point $\zeta_j - 2s = 0$, \ie\ the coupling is marginal.
The result of Disertori et al.\ (which we reproduce at one-loop) means that we can extend this latter result to all loops. Let us now return to considering what happens if this in fact holds nonperturbatively. By making the choice $\alpha_{\hat{\lambda}}(s) = \delta(s)$, we see that
\be
	\lim_{\Lambda \rightarrow \infty} \Lambda
	\der{\hat{\lambda}}{\Lambda} = 0,
	\qquad
	\Rightarrow
	\qquad
	\lim_{\Lambda \rightarrow \infty}
	\Lambda
	\left.\pder{\hat{\lambda}(t,\thetabar)}{\Lambda}\right\vert_{\thetabar} = 0.
\label{eq:imply}
\ee
This is certainly necessary for~\eq{floating-criterion}, but obviously not sufficient. This allows for the
following scenarios:
\begin{enumerate}
	\item There exists a line of inequivalent floating-points which \emph{in the limit $\Lambda 
	\rightarrow \infty$} and only in this limit can be obtained from the Gaussian one via the exactly 
	marginal deformation
	in the $\hat{\lambda}$ direction. Away from this limit, the recipe for generating the floating-points 
	presumably becomes more complicated.
	In this case, the asymptotic safety scenario of Disertori et al.\ holds exactly as originally proposed:
	perturbing the Gaussian floating-point in the said direction in the UV essentially moves us 
	along the line of inequivalent floating-points whereas, for finite $\Lambda$, the various RG 
	trajectories
	flow away from their associated floating-point.
	
	\item It turns out that~\eq{imply} holds at \emph{all} scales. From the perspective of the result of 
	Disertori et al.\ (and our own calculations) this seems unlikely, since the flow of $\lambda$ only
	vanishes in the $\Lambda \rightarrow \infty$ limit. However, it is conceivable that, were we to find a 
	norm that works at all scales, we might find that the flow of  $\hat{\lambda}$---\ie\ the `right' 
	coupling---is actually zero at all scales ($\lambda$ and $\hat{\lambda}$ are the same in the 
	large-$\Lambda$ limit but could turn out to be different for finite $\Lambda$). 
	If this turned out to be the case, then this would mean that 
	one
	could obtain a line of inequivalent floating-points by deforming the Gaussian floating-point in the 
	usual way \emph{at all scales}. Then, the asymptotic safety scenario of Disertori
	et al.\ would be replaced by a scenario in which there is still a non-trivial, nonperturbatively 
	renormalizable theory, but its only scale dependence would be through $\thetabar$.
	(If all this were in the context of a commutative theory, then this option would correspond to
	having found a line of non-trivial conformal field theories, whereas the above option would 
	correspond to having found renormalizable trajectories emanating from a line of conformal
	field theories.)
\end{enumerate}

Let us conclude by saying that, even if $\hat{\lambda}$ turns out to be nonperturbatively
irrelevant, this does not rule out some \emph{other} asymptotic safety scenario. This would
require some non-trivial floating-point (which is not obtainable by a deformation of the Gaussian one,
at any scale) which supports a renormalized trajectory along which the action flows towards the Gaussian theory in the IR. We leave the issues of whether non-trivial floating-points exist, and the renormalized trajectories that they support, for the future.

The rest of this paper is arranged as follows. In \sect{Flow} we introduce the flow equation
that we will use, discuss the form of the effective action and describe the diagrammatics we will
employ to facilitate our computation of the $\beta$-function. \Sect{WR} begins with a review of
Wilsonian renormalization in the commutative setting, in \sect{comm}. Following this,
we generalize the analysis to the noncommutative case in \sect{noncomm}, starting in \sect{Float} where we provide further details about floating-points. We begin \sect{NCRTs} by discussing the renormalized trajectories emanating from floating-points. With this in mind, we classify the eigenperturbations of the Gaussian floating-point according to whether they are relevant, irrelevant or marginal. Just as in the commutative case there is a marginal four-point term at leading order in perturbations about the Gaussian floating-point. Consequently, we go beyond leading order in \sect{beta}, where we present a one-loop computation of the $\beta$-function and discuss how this maps on to existing results.

There are two substantial appendices. In \app{base} we review the matrix base and then discuss
the general form of the Wilsonian effective action. \App{norms} is devoted to computing the norms
of the eigenperturbations of the Gaussian floating-point, which is instrumental in allowing us to correctly classify (ir)relevance.

\section{The Flow Equation}
\label{sec:Flow}

\subsection{The Matrix Polchinski Equation and Variants}

A crucial ingredient of the analysis by Grosse \& Wulkenhaar~\cite{G+W-PC,G+W-2D,G+W-4D} is a flow equation, formulated in the matrix base. In this section, we will review this construction---which is just the matrix version of the Polchinski equation~\cite{Pol}---and then describe a generalization that puts the flow equation in a more convenient form for our subsequent discussion of Wilsonian renormalization.

The central element of any ERG equation is a UV cutoff, which suppresses modes above the effective scale, $\Lambda$.
This can be implemented by modifying the bare 
propagator [\ie\ the inverse of~\eq{CTP}] according to
\be
	\Delta'_{mn;kl} = \frac{\theta}{4(2+m+n)} \delta_{ml} \delta_{nk}
	c(m,n;\thetabar),
\label{eq:EffProp}
\ee
where $c$ is an ultraviolet cutoff function, dying off rapidly if either $ m,n> \thetabar$, and
for which $c(0,0;\thetabar) = 1$ (the reason for the prime, which does not serve to indicate that the propagator has been regularized, will become clear shortly). One sensible choice
for $c$, which we will employ, is
\be
	c(m,n;\thetabar) = K(m/\thetabar) K(n/\thetabar),
\label{eq:sensible}
\ee
where the $K$s are new UV cutoff functions. The cutoff function chosen by Grosse and Wulkenhaar~\cite{G+W-4D} fits into this class. The UV regularized propagator is often referred to  as an
effective propagator.

Mirroring the modification of the propagator, we also modify the kinetic term, such that it is now the
inverse of the effective propagator:
\be
	\hS \equiv  \frac{\nu_4}{2} \phi  \knl{\Delta'^{-1}} \phi.
\label{eq:kinetic}
\ee

To formulate the flow equation, it is convenient to split the full action into the kinetic term and an interaction piece, and so we write
\be
	S = \hS[\phi] + \Sint[\phi],
\label{eq:full}
\ee
where we point out that our definition of the action is such that it includes the volume factor $\nu_4$.

Note that we are free to include two-point vertices in $\Sint$
and so, although the effective propagator looks like it corresponds to a massless theory at
the self-dual point, this is not necessarily true of the full theory. In other words, the splitting between
$\Delta'^{-1}$ and the two-point part of $\Sint$ is down to choice, with~\eq{EffProp} being particularly convenient.

Up to a vacuum energy term, which is uninteresting for the current purposes, the matrix version of the Polchinski equation reads
\be
	-\flow S = 
	\frac{1}{2\nu_4}
	\left(
		\pder{S}{\phi} \knl{\dd'} \pder{\Sigma}{\phi} - \pder{}{\phi} \knl{\dd'} \pder{\Sigma}{\phi}
	\right),
\label{eq:Matrix-Pol}
\ee
where the partial derivative is performed at constant field. The ERG kernel, $\dd'$, is given by the flow of the effective propagator:
\be
	\dot{X} \equiv -\Lambda \der{X}{\Lambda},
\label{eq:dot}
\ee
and we define
\be
	\Sigma \equiv S - 2\hS.
\label{eq:Sigma}
\ee

By direct substitution of~\eq{Sigma} into~\eq{Matrix-Pol}, and by using~\eqs{dot}{kinetic},
it is easy to check that (up to the aforementioned vacuum term) the flow equation can be written just in
terms of $\Sint$, as in Polchinski's original formulation:
\be
	-\flow \Sint = 
	\frac{1}{2\nu_4}
	\left(
		\pder{\Sint}{\phi} \knl{\dd'} \pder{\Sint}{\phi} - \pder{}{\phi} \knl{\dd'} \pder{\Sint}{\phi}
	\right).
\label{eq:Pol}
\ee
This equation is essentially the same as the one in~\cite{G+W-4D}, differing only because we have chosen to incorporate the volume factor in the action. The two terms on the \rhs\ of~\eq{Matrix-Pol} 
or~\eq{Pol} are often called the classical and quantum terms, respectively.

To uncover the noncommutative analogue of fixed-points (\ie\ floating-points), it is convenient to rescale to dimensionless
variables:
\be
	\phi \rightarrow \phi \sqrt{Z} \Lambda,
\label{eq:rescale}
\ee
where $Z$ is the field strength renormalization, and the single power of $\Lambda$ takes care
of the canonical dimension of the field which, just as for a commutative scalar field in four dimensions, is unity. Unfortunately, this rescaling introduces an annoying factor of $1/Z$ on
the \rhs\ of the flow equation. To get rid of these factors, we exploit the huge freedom in formulating ERGs, present as a consequence of the corresponding freedom in the way in which high energy modes are integrated out.

General ERGs are defined according to~\cite{WegnerInv,TRM+JL}:
\be
\label{eq:blocked}
	-\flow e^{-S[\phi]} =  \sum_{m,n} \pder{}{\phi_{mn}} \left(\Psi_{mn}[\phi] e^{-S[\phi]}\right).
\ee
The total derivative on
the \rhs\ ensures that the 
partition function $Z = \int \measure{\phi} e^{-S}$
is invariant under the flow---a fundamental ingredient of any
ERG equation. The functional, $\Psi$, parametrizes (the matrix version of)
a
general Kadanoff blocking procedure~\cite{Kadanoff} 
and so there is considerable choice in its precise
from. We will focus on those blockings for which
\be
\label{eq:Psi}
	\Psi_{mn} = \hf \sum_{kl} \dd^{\mathrm{new}}_{mn;kl} \fder{\Sigma}{\phi_{kl}},
\ee
which clearly reproduces the Polchinski equation, if we identify $\dd^{\mathrm{new}}$
with $\dd'$.

However, rather than making this identification, we
choose $\dd^{\mathrm{new}} = Z \dd$
since now the ERG equation \emph{after the rescaling}~\eq{rescale} is:
\be
	\left(
	 	\partial_t	+ \frac{2+\eta}{2} \phi \cdot \pder{}{\phi}
	 \right)
	 S
	 = 
	\frac{1}{2\chibar_4}
	\left(
		\pder{S}{\phi} \knl{\dd} \pder{\Sigma}{\phi} - \pder{}{\phi} \knl{\dd} \pder{\Sigma}{\phi}
	\right),
\label{eq:Flow}
\ee
where, as before,  $t \equiv \ln \mu /\Lambda$ is the RG-time  and we have defined
\be
	\chibar_4 \equiv 4 \pi^2 \thetabar.
\ee
The reason that $\chibar_4$ appears, and not $\nubar_4 \equiv \nu_4 \Lambda^4$ is that we have
cancelled out the overall power of $\theta$ in $\Delta'$ [see~\eq{EffProp}].
Consequently, we take the unprimed $\Delta$ to be given by
\be
	\Delta_{mn;kl} = \frac{1}{4(2+m+n)} \delta_{ml} \delta_{nk}
	c(m,n;\thetabar).
\label{eq:Rescaled-EP}
\ee
 As usual, $\eta$ is the anomalous dimension of the field, defined by
 \be
 	\eta \equiv \Lambda \der{\ln Z}{\Lambda}.
 \label{eq:eta}
 \ee
The flow equation~\eq{Flow} is the matrix version of the one first written down by Ball et al.~\cite{Ball}.

As an aside, we note that the freedom in $\Psi$ means that we could, if we wished, furnish $\hS[\phi]$ with
interactions. Such an action, which partially parametrizes the residual blocking freedom left over, given the
choices~\eqs{Psi}{Sigma} is called the `seed action'~\cite{aprop} (the rest of the blocking freedom is encoded in our choice of cutoff function). There is little point in introducing a general seed action here, because this makes the flow equation unnecessarily complicated; however, it should be borne in mind that this is necessary when constructing Polchinski-like flow equations for gauge theories~\cite{qcd,qed}.

Before moving on, we will introduce one other version of the flow equation which will be particularly convenient for computing the $\beta$-function. This can be directly obtained from the last flow equation by a change of variables. Rather than rescaling the field by the full scaling dimension, as in~\eq{rescale}, we remove only the anomalous part and then perform an additional rescaling using the four-point coupling, $\hat{\lambda}$. Thus~\eq{rescale} is replaced with
\be
	\phi \rightarrow \phi \sqrt{Z} /\sqrt{\hat{\lambda}}.
\label{eq:rescale-lambdahat}
\ee
This rescaling with the coupling ensures that $1/\hat{\lambda}$ now appears in front of the action and so the expansion in terms of $\hat{\lambda}$ coincides with the expansion in $\hbar$. The flow equation reads:
\be
	\left(-\flow + \frac{\gamma}{2} \phi \cdot \pder{}{\phi} \right)S[\phi] = 
	\frac{1}{2\chi_4}
	\left(
		\classical{S}{\dd}{\Sigma_{\hat{\lambda}}} - \quantum{\dd}{\Sigma_{\hat{\lambda}}}
	\right),
\label{eq:beta-flow}
\ee
where
\begin{align}
	\Sigma_{\hat{\lambda}} & \equiv \hat{\lambda} (S-2\hS),
\\
	\gamma & = \eta- \frac{\beta}{\hat{\lambda}}.
\label{eq:gamma-defn}
\end{align}

\subsection{The Effective Action}
\label{sec:Eff-Ac}

Let us now return to the question of what constraints we place on the actions which populate theory space. First, we choose to consider only those actions which are invariant under 
$\phi \rightarrow -\phi$. To understand the other constraint, it is useful to turn back to the commutative theory, for means of comparison.

As mentioned in the introduction, we demand that actions in the commutative theory are 
`quasi-local'~\cite{ym} meaning that they can be expanded to all orders in powers of derivatives, equivalently powers of momenta. 
In order that this property is preserved by the flow (at least for $\Lambda >0$) we simply need to choose
a quasi-local cutoff function. It will be instructive, for when we go back to the noncommutative case, to see how this ensures that quasi-locality is preserved along the flow, in position space. To this end, consider a derivative expansion of the two-point action:
\be
	S_{\mathrm{2pt}} = 
	\FourInt{x} 
	\left[
		\hf \phi^2(x) + a \phi(x) \partial^2 \phi(x) +\ldots
	\right],
\label{eq:2pt}
\ee
and now consider the effect of the classical term in the flow equation when acting on this:
\be
	\classical{S_{\mathrm{2pt}}}{\dd}{S_{\mathrm{2pt}}}
	=
	\FourInt{x}\FourInt{y}
	\phi(x) \dd(x,y) \phi(y) + \ldots,
\label{eq:2ptx2pt}
\ee
where the ellipsis denotes terms coming from the higher derivative pieces of $S_{\mathrm{2pt}}$.
If we now expand
\[
	\phi(y) = \phi(x) + (y-x)_\mu \pder{\phi}{x_\mu} + \ldots,
\]
and similarly with $\dd(x,y)$, we see that the \rhs\ of~\eq{2ptx2pt} can be rewritten in the form of~\eq{2pt}.
The presence of the cutoff function in the kernel $\dd(x,y)$ is crucial, since it is this that renders the
definite integral over $y$ finite. Indeed, it is clear that
\[
	\FourInt{x} \phi(x)  \FourInt{y} \phi(y) 
\]
cannot be rewritten in the form~\eq{2ptx2pt}. But, if we do not include terms like this at the start, they
are never generated.

Let us now return to the noncommutative case. In the $\star$-basis, we might expect that an effective
action can be written in terms of a single integral, but where now we allow fields to be hit by star multiplications with $\tilde{x}_\mu$:
\begin{multline*}
	S_{\mathrm{eff}} = 
	\FourInt{x}
	\biggl[
		A_0 \phi \star \phi
		+ A_2 \phi \star \tilde{x}_\mu \star \tilde{x}^\mu \star \phi
		+ A_4 \phi \star \tilde{x}_\mu  \star \phi\star \tilde{x}^\mu
		+ \mbox{more two-point terms}
\\
		+ B_0 \phi \star \phi \star \phi \star \phi
		+ \mbox{more four-point terms}
		+\mbox{higher-point terms}
	\biggr]
\end{multline*}

In the matrix base, this translates into effective actions built from a \emph{single} trace containing
$\phi_{mn}$s and $\bigl(\tilde{X}_\mu\bigr)_{kl}$s. Now, in direct analogy with the commutative case,
so long as we choose a kernel $\dd_{mn;kl}$ that can be expanded in terms of $\tilde{X}_\mu$s,
and so long as this kernel properly incorporates a cutoff function, then the single trace structure is
preserved. Of course, this single trace structure will not necessarily be manifest---just as the expression on the \rhs\ of~\eq{2ptx2pt} is not manifestly a single integral term; but all terms apparently containing two or more traces can be rewritten as single integral terms.

In \app{Action-basis} we show both that single trace terms built out of $\phi_{mn}$s and $\bigl(\tilde{X}_\mu\bigr)_{kl}$s provide a good basis and that there is no problem building $\dd_{mn;kl}$ out of
$\tilde{X}$s.

\subsection{Diagrammatics}

Expanding the Wilsonian effective action in powers of the field, the flow equation has a natural
diagrammatic interpretation, consisting of ribbon graphs. To see how this comes about, we first
introduce a diagrammatic expansion for the action:
\be
	S = \hf \phi_{m_1n_1} \phi_{m_2 n_2} \ensuremath{\begin{array}{c}\input{pstex/S_2.pstex_t} \end{array}} 
	+ \frac{1}{4!} \phi_{m_1n_1} \phi_{m_2 n_2} \phi_{m_3n_3} \phi_{m_4 n_4} \ensuremath{\begin{array}{c}\input{pstex/S_4.pstex_t} \end{array}} +\ldots
\label{eq:diag-action}
\ee
There are a number of points to make about these diagrams. First, the lobes---inside each of which we have written $S$---are just there for convenience, since it provides a nice place to put labels. (For example, if we wanted to write down a diagrammatic expansion for the seed action, we could replace the $S$ above with $\hS$.) Thus, once can obtain more familiar-looking diagrams by shrinking the lobes down to a point. Secondly, at this stage, the only expansion that has been performed is the one about vanishing field; in particular, no perturbative expansion of the vertices has been performed. Indeed, if one now substitutes the diagrammatic expansion~\eq{diag-action} into the flow equation, then one can
write down an infinite tower of coupled equations for the vertices. Whilst one can certainly solve this tower perturbatively, order by order---as we will do later---the full solution to these equations contains nonpertubative information, too.

The presence of the lobes will prove particularly useful when we come to do perturbation theory.
As a result of our rescaling of the field with $1/\sqrt{\hat{\lambda}}$, the action has the perturbative expansion
\be
	S \sim \sum_{i=0}^{\infty} \hat{\lambda}^{i-1} S_i.
\ee
Thus we write the perturbative expansion of the $M$-point vertex as:
\[
	\ensuremath{\begin{array}{c}\input{pstex/S_M.pstex_t} \end{array}} \sim 	\frac{1}{\hat{\lambda}} \ensuremath{\begin{array}{c}\input{pstex/S_M_0.pstex_t} \end{array}} + \ensuremath{\begin{array}{c}\input{pstex/S_M_1.pstex_t} \end{array}} + \ldots
\]

\section{Wilsonian Renormalization}
\label{sec:WR}

The reason for performing the rescaling to dimensionless variables using $ \Lambda$
(rather than $\theta$) is
to enable us to define nonperturbatively renormalizable theories, in a particularly simple way.
By renormalizable in the nonperturbative sense, we mean the following: having integrated out
degrees of freedom between the bare scale and the effective scale we want to know if there
are any theories for which the limit
\[
	\lim_{\Lambda_0 \rightarrow \infty} S_{\Lambda,\Lambda_0}[\phi]
\]
can be safely taken in the sense that any divergences can be absorbed into a finite number of
parameters. To understand which theories satisfy this criterion, we will first review the Wilsonian
picture of renormalization in commutative theories.

\subsection{Commutative Theories}
\label{sec:comm}

In the commutative setting, as discussed in the introduction, one class of theories which are nonperturbatively renormalizable are
conformal field theories: since they are, by definition, scale independent, they are trivially independent
of $\Lambda_0$ which can thus clearly be sent to infinity without any difficulty. From the perspective
of the ERG, conformal theories follow simply from \emph{fixed-points}:
\be
	\partial_t S_*[\varphi] = 0.
\label{eq:Comm-FP}
\ee
The point of the rescalings is that, by measuring all dimensionful quantities in terms of $\Lambda$
(momenta, too, should be rescaled), independence of $\Lambda$ (equivalently $t$) implies
independence of all scales.

Once the critical fixed-points have been identified in a commutative field theory, scale \emph{dependent}
renormalizable theories can be found by considering the aptly named `renormalized trajectories' emanating from the fixed-points~\cite{Wilson}. 
These correspond to perturbing the fixed-point action in the
directions relevant \emph{\wrt\ this fixed-point}. Assuming for the sake of simplicity that there are no
marginally relevant operators,\footnote{Marginally relevant operators can be included by adding terms which, as $\Lambda \rightarrow \infty$, sink back into the fixed-point only as a power of $t$ (\ie\ logarithmically with $\Lambda$).} the boundary condition for such a trajectory is~\cite{TRM-Elements}:
\be
	\lim_{\Lambda \rightarrow \infty} S_t[\varphi] = 
	S_*[\varphi] +
	\sum_{i=1}^n \alpha_i e^{\critexp_i t} \mathcal{O}_i[\varphi],
\label{eq:Comm-RT}
\ee
where the sum is over the $n$ relevant directions, spanned by the operators $\mathcal{O}_i$, the $\alpha_i$ are constants, and the $\critexp_i >0$
are the RG eigenvalues.\footnote{$\hat{\lambda}$ without an index a coupling, not to
be confused with the RG-eigenvalues, $\critexp_i$.} 
 The renormalized couplings, $g_i$, (which include the rescaled mass, if appropriate) are conjugate to the operators, in the sense that
\be
	\lim_{\Lambda \rightarrow \infty} g_i \sim \alpha_i e^{\critexp_i t}.
\ee
It is straightforward to demonstrate the renormalizability of such trajectories by noting that,
at the effective scale, dependence on $\Lambda$ and $\alpha_i$ can be traded for the
renormalized couplings and anomalous dimension (see~\cite{TRM-Elements} for a very simple proof):
\be
	S_t[\varphi] (\alpha_i)= S[\varphi] (g_i(t),\eta(t)).
\ee
The \rhs\ side of this equation is in so-called `self-similar form' meaning that all scale dependence
occurs only through the renormalized couplings and anomalous dimension. In particular,
there is no explicit dependence on $\Lambda/\Lambda_0$, and so the theory is renormalizable.

In four dimensions it is worth pausing in order to understand how to reconcile the lack of self-similar trajectories, and
hence the triviality of the theory, with perturbative renormalizability of $\lambda \varphi^4$ theory.
Perturbatively,
it is true that one can write down a self-similar action in terms of $m(t), \lambda(t)$
and $\eta(t)$. However, the various perturbative series are
ill-defined, as a consequence of ultraviolet renormalons. To rectify this problem necessitates the
presence of terms depending on $\Lambda/\Lambda_0$; a new scale has been introduced, and
so self-similarity is destroyed.

In a different way of looking at things, these renormalons can be eliminated by passing to the ``effective expansion''.
This effective expansion is written not in terms of only one coupling constant but in terms of an entire series of effective coupling constants associated to a ladder of intermediate scales between $\Lambda_0$ and the IR. Although the new series does not possess renormalons it violates self-similarity due to the presence of the said intermediate scales (for a detailed introduction in multiscale analysis and the effective expansion see~\cite{Riv-book}).

\subsection{Noncommutative Theories} 
\label{sec:noncomm}

\subsubsection{Floating-Points}
\label{sec:Float}

As discussed in \sect{renorm}, rather than searching for fixed-points to classify nonperturbatively renormalizable theories, we instead consider \emph{floating-points} defined according to
\be
	\left.\partial_t\right|_{\phi, \thetabar} S_* [\phi] = 0,
\label{eq:floating}
\ee
(we henceforth take $*$ to exclusively denote floating-points).
This criterion clearly excludes dependence on $\theta \Lambda_0^2$, since
\[
	\theta \Lambda_0^2 = \thetabar \frac{\Lambda_0^2}{\Lambda^2}.
\]

As a straightforward example of this, let us confirm that what we might have called the Gaussian fixed-point is, when written in the matrix base using rescaled variables, the Gaussian floating-point:
\be
	\frac{\nu_4}{2} \phi \knl{\Delta'^{-1}}\phi \rightarrow 
	\frac{\nubar_4}{2} \phi \knl{\frac{\Delta^{-1}}{\thetabar}} \phi,
\label{eq:GFP}
\ee
where the arrow indicates the rescaling~\eq{rescale} (with, of course, $Z=1$). Clearly, the action
does not satisfy the fixed-point condition, since $\left.\partial_t\right|_\phi S \neq 0$ (this is true even before rescaling, as a consequence of the $\Lambda$ buried in the cutoff function), but it does satisfy~\eq{floating}.

Before moving on, it is worth pointing out 
that there is in fact not a single Gaussian floating-point but
a \emph{line of equivalent floating-points}, each linked to the rest by a reparametrization of the
field (this is analogous to the commutative case~\cite{TRM-Elements}). This can be straightforwardly checked by  substituting the ans\"{a}tz
\[
	\Sint_* = \chibar_4 \hf \phi \knl{Y(\thetabar)} \phi
\]
into~\eq{Flow} (with $\eta_* = 0$), upon which it is found that the general Gaussian floating-point solution takes the form
\[
	S_* = \hf \phi \knl{(\one - W^{-1} \cdot  \Delta)^{-1} \cdot \Delta^{-1}} \phi,
\]
where $W$ is independent of $\thetabar$, and arises as an integration constant.

The floating-points equivalent to $\Sint=0$ are those for which can write
\[
	S_* = \hf \sum_{m,n,k,l}
	\phi_{mn} F_{mn;kl} \phi_{kl}, 
	\qquad
	\mathrm{with \ }
	F_{mn;kl} 
	= \big[4(2+m+n) + \ldots\big]\delta_{ml}\delta_{nk}, 
\]
where the ellipsis denotes terms higher order in the indices $m$ and $n$. In other words, if the only effect of $W$ is only to modify the non-universal part of the two-point function corresponding to the (inverse) cutoff function, then we are dealing with a floating-point equivalent to the one with $\Sint=0$. Contrariwise, if the effect of $W$ is to produce an action which is, say, away from the self-dual theory, then this floating-point is not equivalent to the one we are interested in.

\subsubsection{Renormalized Trajectories}
\label{sec:NCRTs}

Now that we have found the noncommutative analogue of fixed-points, we should attempt to 
define the associated renormalized trajectories. This requires that we understand what
we mean by (ir)relevant operators in the noncommutative context and we shall start
by using the Gaussian floating-point as a test case.

As it happens, there is a major subtlety in the matrix base that is perfectly illustrated by perturbing the
Gaussian floating-point by a mass:
\be
	S_t = 
	 \frac{\nubar_4}{2}
	\left(
		\phi \cdot \tilde{X}_\mu \cdot \tilde{X}^\mu \cdot \phi
		+
		\frac{M^2}{\Lambda^2} \phi \cdot \phi
		+\ldots
	\right),
\ee
where the ellipsis represents the regulator contributions, which we do not need for this argument.

Now, at first sight, the mass is relevant (as should be expected) since, taking $\Lambda \rightarrow \infty$, the action
sinks back into the floating-point solution. It is very important to
note that we do not hold $\thetabar$ constant when taking this limit: $\thetabar$ is only held constant for the purpose of finding floating-points; once found, we are interested in how perturbations evolve with a change of scale without any constraints. However, we know from~\eq{tildeX} that buried in each $\tilde{X}$ is a $1/\sqrt{\thetabar}$. From this perspective, the behaviour of the kinetic term and the mass term are the same as we take the limit $\Lambda \rightarrow \infty$, suggesting that the mass is only marginal. This conclusion, were it to hold true, would be very surprising. We will see below how to resolve this conundrum.

To classify the eigenperturbations, we perturb the floating-point action:
\be
	S_t[\phi](\thetabar) = S_*[\phi](\thetabar) + T_t[\phi](\thetabar)
\label{eq:NC-linearize}
\ee
where, in what follows, we will consider $T$ to be small.

Our aim is to substitute~\eq{NC-linearize} into the flow equation and to start by working at linear order in $T$.
To make life easier, we will
anticipate the form of $T$ at this order and so write
\be
	T_t[\phi](\thetabar) \sim \nubar_4 \sum_i \alpha_i e^{\zeta_i t}
	\mathcal{Q}_i[\phi](\thetabar),
\label{eq:T-linear}
\ee
where the sum over $i$ is over all operators. In fact,  as it stands, the sum over $i$ is counting separately identical operators multiplied by different powers of $\thetabar$; we will rectify this below.

We now substitute~\eq{NC-linearize} into our flow equation~\eq{Flow}, utilizing~\eq{T-linear}, and linearize:
\be
	\left(
		\zeta_i -4 - 2 \thetaflow + \phi \cdot \pder{}{\phi}
	\right)\mathcal{Q}_i
	=
	-\frac{1}{2\chibar_4}
	\pder{}{\phi} \knl{\dd} \pder{\mathcal{Q}_i}{\phi}.
\label{eq:flow-O}
\ee
Defining
\be
	\mathcal{Q}'_i \equiv
	\exp
	\left(
		\frac{1}{2\chibar_4} \pder{}{\phi} \knl{\Delta} \pder{}{\phi}
	\right) \mathcal{Q}_i,
\ee
we see that~\eq{flow-O} implies:
\be
	\left(
		\zeta_i -4 - 2 \thetaflow + \phi \cdot \pder{}{\phi}
	\right)\mathcal{Q}'_i
	= 0,
\label{eq:flow-O'}
\ee
where we have used
\[
	\left[
		\frac{1}{2\chibar_4} \pder{}{\phi} \knl{\Delta} \pder{}{\phi}, \ \phi \cdot \pder{}{\phi} 
	\right]
	=
	\frac{1}{\chibar_4}  \pder{}{\phi} \knl{\Delta} \pder{}{\phi}.
\]
We will suppose that the physically interesting solutions to~\eq{flow-O'} are those for which
$\mathcal{Q}'_i$ is a homogenous polynomial in $\phi$.
Recalling from \sect{Eff-Ac} that we can take the $\mathcal{Q}'_i$ to be single trace terms
built out of $\tilde{X}$s and $\phi$s 
we write
\be
\begin{split}
	\mathcal{Q}'_i 	
	& = 
	\thetabar^{s}
	\mathcal{Q}'^{(\varpi)}_{\mu_1 a_1b_1, \ldots, \mu_{2\numX} a_{2\numX} b_{2\numX}; c_1 d_1,			\ldots, c_J d_J}
	\tilde{X}^{\mu_1}_{a_1b_1} \cdots \tilde{X}^{\mu_{2\numX}}_{a_{2\numX} b_{2\numX}}
	\phi_{c_1 d_1}\cdots \phi_{c_J d_J},
\\
	& = 
	\thetabar^{s}
	u^{(\varpi,\numX)}_{m_1 n_1 \cdots m_J n_J}
	\phi_{m_1 n_1}\cdots \phi_{m_J n_J},
\label{eq:Q'-Decomp}
\end{split}
\ee
where the label, $i$,
includes the following:
\begin{enumerate}
	\item the number of fields, $J$, which we take to be a positive (even) integer;
	
	\item the number of $\tilde{X}_\mu$s, which we denote by $2\numX$;
	
	\item the number of powers of $\thetabar$ \emph{in addition} to the $-\numX$ associated
	with the $2\numX$ $\tilde{X}$s, which we denote by $s$;

	\item an additional index, $\varpi$, which runs over the number of independent operators
	built out of the above ingredients.
\end{enumerate}

Substituting into~\eq{flow-O'} yields
\be
	\zeta_i - 2(s-\numX) = 4 - n.
\label{eq:condition}
\ee

We now observe an obvious but crucial point: operators $\mathcal{Q}'_i$ 
with different $\thetabar$ dependence but
with the same number of $\tilde{X}$s and the same number of $\phi$s, tied together in the same way, are the same operator. 
To illustrate this, consider the following contributions to the $\phi \cdot \phi$ term:
\[
	\hf a \phi \cdot \phi, \qquad \hf \frac{c}{\thetabar} \phi \cdot \phi, \ldots.
\]
The point is that we \emph{define}\footnote{Actually, there is a different definition of the mass available, as we will discuss shortly, but this is neither here nor there for this discussion.} the mass squared to be the total coupling in front of the $1/2 \phi \cdot \phi$ term.
Thus, if both the above terms were present, we would define the (dimensionless) mass to be $M^2 = a + c/\thetabar$.

Consequently, we rewrite~\eq{Q'-Decomp} as
\be
	\mathcal{Q}'_i 	\equiv \thetabar^{s}\mathcal{O}'_j,
\label{eq:Q'-O'}
\ee
where $j$ stands for just the triplet $(J,\numX,\varpi)$, and the $\mathcal{O}'_j$ depend
on $\thetabar$ only via instances of $\tilde{X}$.
From this, it follows that we can rewrite~\eq{NC-linearize} as
\be
	S_t[\phi] = S_*[\phi]  + \nubar_4 \sum_j \int \! ds \, \alpha_j(s) e^{\zeta_j t}\thetabar^s 
	\mathcal{O}_j[\phi],
\label{eq:Pert-Final}
\ee
where~\eq{condition} is understood and
\be
	\mathcal{O}_j = 
	\exp
	\left(
		- \frac{1}{2\chibar_4} \pder{}{\phi} \knl{\Delta} \pder{}{\phi}
	\right) \mathcal{O}'_j.
\label{eq:O_j}
\ee

At this stage, it is very tempting to identify the  couplings of the operators, in the vicinity of the Gaussian 
floating-point, according to:
\[
	g_j[\alpha](t, \thetabar) \sim \int \! ds \, \alpha_j(s) e^{\zeta_j t} \thetabar^s.
\]
However, this is not quite correct, as we now explain [indeed, one can check using~\eq{condition} that,
with this identification, we find \eg\ that the mass is marginal].

Let us begin by taking a step back. We wish to consider flows in the vicinity of the Gaussian floating-point, and we know that such flows are spanned by the $\mathcal{O}_j$. Now, given some generic
action, one question we will want to ask is how this action decomposes onto the $\mathcal{O}_j$. Clearly, the result of this depends on the norm (to be defined below) of the $\mathcal{O}_j$. Constant contributions to the norm are neither here nor there: it is a matter of convention whether we place these
constants in the operators or their couplings. However, we will find that, in general, the norm also has
components which depend on $\thetabar$, and these are crucial. Defining $\hat{\mathcal{O}}_j$ such
that they have constant norm, we can rewrite~\eq{Pert-Final} as
\be
	S_t[\phi]
	 = S_*[\phi]  + \nubar_4 \sum_j \int \! ds \, \alpha_j(s) e^{\zeta_j t}\thetabar^s 
	\norm{\mathcal{O}_j}
	\hat{\mathcal{O}}_j[\phi],
\label{eq:Pert-Final-b}
\ee
where we should identify the couplings in the vicinity of the floating-point as
\be
	\hat{g}_j[\alpha](t, \thetabar) \sim \int \! ds \, \alpha_j(s) e^{\zeta_j t} \thetabar^s
	\norm{\mathcal{O}_j}.
\label{eq:couplings}
\ee
Clearly, $\thetabar$ dependence of the norm affects the $\Lambda$ dependence of the couplings.

We will define a norm by 
recognizing from~\eq{O_j} that the $\mathcal{O}$s are somewhat reminiscent of Hermite polynomials.
Thus,  we tentatively define an inner product according to:
\be
	\inner{a}{b} \stackrel{?}{\equiv}
	\frac{
		\ds
		\int \mathcal{D} \phi e^{-\frac{\chibar_4}{2} \phi \cdot \Delta^{-1} \cdot \phi} ab 
	}
		{
		\ds
		\int \mathcal{D} \phi e^{-\frac{\chibar_4}{2} \phi \cdot \Delta^{-1} \cdot \phi}
	},
\label{eq:product-1}
\ee
with the norm given by
\be
	\norm{a} = \sqrt{\inner{a}{a}}.
\ee
As we will see in a moment, this definition of the inner product will have to be tweaked, but it is a good
place to start.

Let us now investigate the inner product between a pair of $\mathcal{O}$s both with $J=2$. Making the $J,\ \numX$ and $\varpi$ represented by $j$ manifest, and using~\eq{O_j}, we have:
\be
	\mathcal{O}^{(\varpi,\numX)}_2 = 
	\exp
	\left(
		- \frac{1}{2\chibar_4} \pder{}{\phi} \knl{\Delta} \pder{}{\phi}
	\right)
	\mathcal{O}'^{(\varpi,\numX)}_{mnkl}\phi_{mn} \phi_{kl}
\ee
and so
\be
	\mathcal{O}^{(\varpi,\numX)}_2 = 
	u^{(\varpi)}_{mnkl}
	\left(
		\phi_{mn} \phi_{kl}
		-\frac{\Delta_{mnkl}}{\chibar_4}
	\right).
\ee
It directly follows that\footnote{Whilst $\mathcal{O}$s with different values of $J$ are orthogonal, we do not find that \eg\ the mass operator and the kinetic operator are orthogonal. Of course, we could construct an orthogonal basis, but there is no need to do so.}
\be
	\inner{\mathcal{O}^{(\varpi,\numX)}_2}{\mathcal{O}^{(\varpi',\numX')}_2}
	=
	\frac{1}{\chibar^2_4}
	u^{(\varpi,\numX)}_{mnkl}  u^{(\varpi',\numX')}_{yzwx}
	\left(
		\Delta_{mn;yz} \Delta_{kl;wx} + \Delta_{mn;wx} \Delta_{kl;yz}
	\right).
\ee

Taking $\mathcal{O}^{(\mathrm{K})}_2$ to be the self-dual kinetic operator we have:
\be
	u^{(\mathrm{K})}_{mnkl} = 
	\frac{2(m+n+2)}{\thetabar} \delta_{nk} \delta_{ml},
\ee
from which it follows that
\be
	\norm{\mathcal{O}^{(\mathrm{K})}_2 }^2
	=
	\frac{1}{2(2\pi)^4 \thetabar^4} \sum_{m^1,n^1,m^2,n^2} c(m^1,n^1,m^2,n^2;\thetabar). 
\ee
Immediately, we see that the result of the sum will depend on $\Lambda$. 

We will now place the additional requirement on our inner product that the redundant operator corresponding to reparametrizations of the field that maps us between equivalent realizations of
our theory comes out as exactly marginal, at all scales. (We have already seen an example of such equivalent realizations, when we discussed the line of equivalent Gaussian floating-points in
\sect{Float}.) However, it is only at the Gaussian floating-point that we already know what this operator is (away from here, we would have to compute it), and here it is just $\mathcal{O}^{(\mathrm{K})}$.
Momentarily shutting our eyes, let us go ahead and evaluate
$\norm{\mathcal{O}^{(\mathrm{K})}_2}$ in the large-$\Lambda$ limit.
To leading
order in $\Lambda$ we can replace each sum with an integral and remove the cutoff functions
by taking the upper limits of the integrals to be $\thetabar$. Thus we find that
\be
	\lim_{\Lambda \rightarrow \infty} \norm{\mathcal{O}^{(\mathrm{K})}_2 }^2
	=
	\frac{1}{2(2\pi)^4 \thetabar^4}
	\int_0^{\thetabar} dm^1 dm^2 dn^1 dn^2
	=
	\frac{1}{32\pi^4}.
\ee
Therefore, the kinetic operator, \emph{with the $\thetabar$s from the $\tilde{X}$s included} has constant
norm in the large-$\Lambda$ limit. Moreover, the associated coupling  is
\be
	\lim_{\Lambda \rightarrow \infty}
	\hat{g}^{(\mathrm{K})} \sim \int \! ds \, \alpha^{(\mathrm{K})}(s) e^{\zeta^{(\mathrm{K})} t} \thetabar^{s}.
\ee
But, using~\eq{condition}, and noting that $\numX = 1$, we see that  
$\zeta^{(\mathrm{K})} -2s = 0$. Therefore, the overall $\Lambda$ dependence
vanishes and so we see that the kinetic operator is marginal.

Consequently, we can use our inner product so long as we compute both sides in the limit $\Lambda \rightarrow \infty$, and only ask questions of trajectories which sink into the Gaussian fixed-point, in this limit. Fortunately, this is precisely the scenario we are interested in! 

Thus, for perturbations of the floating-point action, understood to be spawned at $\Lambda \rightarrow \infty$, we can take the following inner product:
\be
	\lim_{\Lambda \rightarrow \infty} \inner{a}{b} =
	\lim_{\Lambda \rightarrow \infty}
	\frac{
		\ds
		\int \mathcal{D} \phi e^{-\frac{\chibar_4}{2} \phi \cdot \Delta^{-1} \cdot \phi} ab 
	}
		{
		\ds
		\int \mathcal{D} \phi e^{-\frac{\chibar_4}{2} \phi \cdot \Delta^{-1} \cdot \phi}
	}
\label{eq:product-2}
\ee

Next let us look at the mass\footnote{
Note that we can define the mass in two ways: either as the coefficient in front of $\mathcal{O}^{(M)}$
or as the coefficient in front of all $\phi\cdot\phi$ terms in the action. This latter definition will pick up contributions from the other $\mathcal{O}$s.
}
operator:
\be
	u^{(M)}_{mnkl} = \delta_{nk} \delta_{ml}.
\ee

From this we find that
\be
	\norm{\mathcal{O}^{(M)}} = \frac{1}{16 \chibar^2_4} \sum_{m^1,n^1,m^2,n^2} 
	\frac{c^2(m^1,n^1,m^2,n^2;\thetabar)}{(2+m^1+n^1+m^2+n^2)^2}. 
\ee
Employing the same methodology as before (this is done in detail in \app{norms}) to pick out the leading behaviour in the $\Lambda \rightarrow \infty$ limit, we find that
\be
	\lim_{\Lambda \rightarrow \infty} \norm{\mathcal{O}^{(M)}} \sim \mathrm{const}.
\ee
Note that, whilst one would na\"{\i}vely expect the leading behaviour to go like $\ln \thetabar$, these
contributions miraculously cancel out.

Consequently, we identify the (dimensionless) mass  squared as:
\be
	\lim_{\Lambda \rightarrow \infty}
	\hat{g}^{(M)} \sim \int \! ds \, \alpha^{(M)}(s) e^{\zeta^{(M)} t} \thetabar^{s}.
\ee
Using~\eq{condition}, we see that
\be
	\lim_{\Lambda \rightarrow \infty} \hat{g}^{(M)} \sim \frac{1}{\Lambda^2},
\ee
and so the mass is relevant---as it should be---with RG eigenvalue $+2$.

The above results pertaining to the kinetic operator and the mass operator lead us to suspect that:
\emph{operators built out of $\tilde{X}$s---with the associated $\thetabar$s included---have
constant norm, with respect to the inner product~\eq{product-2}, in the limit of large $\Lambda$.} This is actually highly non-trivial, due
to the possible appearance of terms proportional to $\ln \thetabar$. However, just as they cancel out
for the mass term, so we have proven in  \app{norms} that they cancel out for all marginal terms as well
as certain families of irrelevant terms. In fact, it is likely that they cancel out in complete generality, but we have not completed the proof of this. Until this is proven, it is possible that certain irrelevant couplings in fact should come with additional $\ln \thetabar$ factors. However, even if such terms do turn out to be present (which we doubt), this will not change the classification of any of the operators.

We are now in a position to classify the (ir)relevance of the various couplings. Let us recall~\eq{condition}:
\[
	\zeta_i - 2(s-\numX) = 4 - n.
\]
By inspection of~\eq{couplings}, we deduce that relevant or marginal couplings satisfy the constraint
\be
	\critexp_j \equiv \zeta_j - 2s  \geq 0 \qquad  \Rightarrow \qquad 4 \geq n+2\numX.
\label{eq:RG-eigen}
\ee
Note that the $\critexp_j$ are the noncommutative analogue of the RG-eigenvalues.
From~\eq{RG-eigen}, we directly uncover the expected result that, at the Gaussian floating-point, the mass squared ($n=2$, $\numX=0$) is
relevant; the kinetic term ($n=2$, $\numX=1$) is marginal (though this something we have required, rather than an independent result);
the deviation of the harmonic oscillator coupling from the self-dual point, $\omega$, ($n=2$, $\numX=1$) is marginal; the four-point
$\mathrm{tr}\, \phi^4$ coupling ($n=4, \numX=0$) is marginal; all other couplings are irrelevant.

Let us comment that this analysis would be spoilt by the inclusion of multi-trace terms. But, as argued in \sect{Eff-Ac}, these can be consistently excluded for the theories we are interested in.

\subsubsection{Beyond Leading Order}
\label{sec:beta}

The final task is to determine whether the marginal couplings are marginally relevant or marginally irrelevant. Actually, we will just ask this question of $\hat{\lambda}$---which amounts to performing a calculation of the $\beta$-function---since the computation will make it apparent that we can directly use earlier results~\cite{G+W-beta}.
Rather than persisting
with writing the action in the eigenoperator basis, we will perform the computation in a more standard way by  identifying $\hat{\lambda}/4!$ as the coupling in front of the total $\phi\cdot\phi\cdot\phi\cdot\phi$ contribution to the action (at least in the large-$\Lambda$ limit, where we know how to compute operator norms). This definition of $\hat{\lambda}$ will, in general, differ from the one where we take it to be the coefficient in front of $\hat{\mathcal{O}}^{(\lambda)}$. However, the definitions
coincide when there are no other operators with four-point pieces which contribute to the action. This is, crucially, precisely the situation we are interested in since we are looking for trajectories which, in the limit $\Lambda\rightarrow\infty$, approach the Gaussian floating-point along the $\hat{\mathcal{O}}^{(\lambda)}$ direction.
This definition of the coupling is both technically easier to work with and also allows our calculation to be transparently compared with
existing work. Indeed, it will become immediately apparent that our one-loop calculation can be
mapped on to the one performed by Grosse \& Wulkenhaar in~\cite{G+W-beta}. With this in mind, we note that Grosse \& Wulkenhaar found that the $\beta$-function is positive (in a limit to be described below), unless the theory is at the self-dual point, whereupon it vanishes. For the sake of simplicity, then, we will directly compute at the self-dual point, since it is for this theory that we stand the best chance of evading triviality.

What we would like to achieve with our calculation is a computation of the flow of $\hat{\lambda}$ for an action which is, to very good approximation, the Grosse \& Wulkenhaar action at the bare scale. In such a computation (as we will see below), we must be careful that there are no hidden running couplings (equivalently, additional scales). But this is actually guaranteed, \emph{at least within perturbation theory}. The point is that
the Grosse \& Wulkenhaar action is perturbatively renormalizable~\cite{G+W-4D}. One way of rephrasing this is to say that all dependence of the action on the effective scale\footnote{Assuming we were to work in dimensionless units by scaling the canonical dimension out of the field, in the usual way.} occurs only through $\hat{\lambda}(t,\thetabar)$ [and, should we take a massive theory, on $m(t,\thetabar)$], and $\thetabar$. Thus, at the perturbative level, we have a self-similar action: there are no hidden scales!
As mentioned in \sect{comm}, beyond perturbation theory self-similarity is violated by the presence of UV renormalons.

To compute the $\beta$-function, we define~\cite{Trivial}
\be
	-\nu_4\dual[\phi] \equiv \ln 
	\left[
		\expoplam
		e^{-\Sint[\phi]}
	\right].
\ee
Substituting this into the flow equation~\eq{beta-flow} we obtain:
\be
	\left[
		\flow + 
		\left(
			\frac{\gamma}{2} + \frac{\beta}{\hat{\lambda}}
		\right)
		\phi \cdot \pder{}{\phi}
	\right]
	\dual[\phi]
	=
	\left(
		\frac{\gamma}{\hat{\lambda}} + \frac{\beta}{\hat{\lambda}^2}
	\right)
	\frac{1}{2\theta}
	\phi \cdot \Delta^{-1} \cdot \phi.
\ee

Defining the field expansion of $\dual$ similarly to that of the action
\be
	\nu_4 \dual[\phi] = 
	\hf \dualv{2}_{m_1n_1m_2n_2} \phi_{m_1n_1}\phi_{m_2n_2}
	+ \frac{1}{4!}
	\dualv{4}_{m_1n_1m_2n_2m_3n_3m_4n_4}
	\phi_{m_1n_1} \phi_{m_2 n_2} \phi_{m_3n_3} \phi_{m_4 n_4}
	+\ldots,
\label{eq:D-flow}
\ee
we need to look at the two-point and four-point flows in order to derive a pair of coupled equations
for $\beta$ and $\gamma$. If we do this at the two-point level, then the leading (\ie\ classical) contribution to the \emph{action} comes from $\hS$; in other words, the two-point pieces of $\Sint$---out of which $\dual$ is built---start at one-loop. Suppressing indices, we therefore write the 
perturbative expansion of $\dualv{2}$ as
\begin{align}
	\hf \dualv{2}_0 & = 0,
\\
	\hf \dualv{2}_1 & = \hf \ensuremath{\begin{array}{c}\begin{picture}(0,0)%
\epsfig{file=pstex/S_2-1-nol.pstex}%
\end{picture}%
\setlength{\unitlength}{3947sp}%
\begingroup\makeatletter\ifx\SetFigFont\undefined%
\gdef\SetFigFont#1#2#3#4#5{%
  \reset@font\fontsize{#1}{#2pt}%
  \fontfamily{#3}\fontseries{#4}\fontshape{#5}%
  \selectfont}%
\fi\endgroup%
\begin{picture}(418,847)(4367,-3372)
\put(4439,-3017){\makebox(0,0)[lb]{\smash{{\SetFigFont{12}{14.4}{\rmdefault}{\mddefault}{\updefault}{\color[rgb]{0,0,0}$\Sint_1$}%
}}}}
\end{picture}%
 \end{array}} - \frac{1}{2\chi_4}
	\left[
		\frac{8}{4!} \ensuremath{\begin{array}{c}\begin{picture}(0,0)%
\epsfig{file=pstex/S_4_0-D.pstex}%
\end{picture}%
\setlength{\unitlength}{3947sp}%
\begingroup\makeatletter\ifx\SetFigFont\undefined%
\gdef\SetFigFont#1#2#3#4#5{%
  \reset@font\fontsize{#1}{#2pt}%
  \fontfamily{#3}\fontseries{#4}\fontshape{#5}%
  \selectfont}%
\fi\endgroup%
\begin{picture}(661,811)(4250,-3270)
\put(4534,-3009){\makebox(0,0)[lb]{\smash{{\SetFigFont{12}{14.4}{\rmdefault}{\mddefault}{\updefault}{\color[rgb]{0,0,0}0}%
}}}}
\end{picture}%
 \end{array}}
		+
		\frac{4}{4!} \ensuremath{\begin{array}{c}\begin{picture}(0,0)%
\epsfig{file=pstex/S_4_0-D-np.pstex}%
\end{picture}%
\setlength{\unitlength}{3947sp}%
\begingroup\makeatletter\ifx\SetFigFont\undefined%
\gdef\SetFigFont#1#2#3#4#5{%
  \reset@font\fontsize{#1}{#2pt}%
  \fontfamily{#3}\fontseries{#4}\fontshape{#5}%
  \selectfont}%
\fi\endgroup%
\begin{picture}(453,958)(4355,-3372)
\put(4534,-3009){\makebox(0,0)[lb]{\smash{{\SetFigFont{12}{14.4}{\rmdefault}{\mddefault}{\updefault}{\color[rgb]{0,0,0}0}%
}}}}
\end{picture}%
 \end{array}}
	\right].
\end{align}
At the two-point level, where there is a difference between $S$ and $\Sint$, we have explicitly indicated
that we are taking contributions only from $\Sint$, as the definition of $\dual$ instructs us.

Introducing the perturbative expansions of $\beta$ and $\gamma$ according to
\be
	\beta \sim \sum_{i=1}^\infty \hat{\lambda}^{i+1} \beta_i, 
	\qquad
	\gamma \sim \sum_{i=1}^\infty \hat{\lambda}^{i} \gamma_i
\ee
and substituting these expansions and our diagrammatic expressions into~\eq{D-flow}, we find that
\be
	\flow
	\left[
		\ensuremath{\begin{array}{c}\begin{picture}(0,0)%
\epsfig{file=pstex/S_2-1-nol.pstex}%
\end{picture}%
\setlength{\unitlength}{3947sp}%
\begingroup\makeatletter\ifx\SetFigFont\undefined%
\gdef\SetFigFont#1#2#3#4#5{%
  \reset@font\fontsize{#1}{#2pt}%
  \fontfamily{#3}\fontseries{#4}\fontshape{#5}%
  \selectfont}%
\fi\endgroup%
\begin{picture}(418,847)(4367,-3372)
\put(4439,-3017){\makebox(0,0)[lb]{\smash{{\SetFigFont{12}{14.4}{\rmdefault}{\mddefault}{\updefault}{\color[rgb]{0,0,0}$\Sint_1$}%
}}}}
\end{picture}%
 \end{array}} - \frac{1}{12\pi^2\theta}  \ensuremath{\begin{array}{c}\begin{picture}(0,0)%
\epsfig{file=pstex/S_4_0-D.pstex}%
\end{picture}%
\setlength{\unitlength}{3947sp}%
\begingroup\makeatletter\ifx\SetFigFont\undefined%
\gdef\SetFigFont#1#2#3#4#5{%
  \reset@font\fontsize{#1}{#2pt}%
  \fontfamily{#3}\fontseries{#4}\fontshape{#5}%
  \selectfont}%
\fi\endgroup%
\begin{picture}(661,811)(4250,-3270)
\put(4534,-3009){\makebox(0,0)[lb]{\smash{{\SetFigFont{12}{14.4}{\rmdefault}{\mddefault}{\updefault}{\color[rgb]{0,0,0}0}%
}}}}
\end{picture}%
 \end{array}} - \frac{1}{24\pi\theta}  \ensuremath{\begin{array}{c}\begin{picture}(0,0)%
\epsfig{file=pstex/S_4_0-D-np.pstex}%
\end{picture}%
\setlength{\unitlength}{3947sp}%
\begingroup\makeatletter\ifx\SetFigFont\undefined%
\gdef\SetFigFont#1#2#3#4#5{%
  \reset@font\fontsize{#1}{#2pt}%
  \fontfamily{#3}\fontseries{#4}\fontshape{#5}%
  \selectfont}%
\fi\endgroup%
\begin{picture}(453,958)(4355,-3372)
\put(4534,-3009){\makebox(0,0)[lb]{\smash{{\SetFigFont{12}{14.4}{\rmdefault}{\mddefault}{\updefault}{\color[rgb]{0,0,0}0}%
}}}}
\end{picture}%
 \end{array}}
	\right]
	=
	(\gamma_1 + \beta_1) \frac{\Delta^{-1}}{\theta}.
\label{eq:2pt-flow-a}
\ee

Now, to convert this expression into an equation relating $\beta_1$ and $\gamma_1$ requires that
we specify a renormalization condition. Since we have removed the field strength renormalization from
the propagator, by means of a field rescaling, we can demand that the kinetic term is canonically normalized. Recalling that
\[
	\Delta^{-1}_{mn;kl} = 4(2+m+n) \delta_{ml}\delta_{nk},
\]
we can insist that the part of $\Sint_{mn;kl}$ linear in the indices vanishes. Therefore, if we 
specialize~\eq{2pt-flow-a} to the piece linear in the indices, then the first term on the \lhs\ can be discarded.

As for the next two diagrams, the first observation we make is that, at the classical level,
the only four-point vertex we have is the one corresponding to $1/4! \Tr \phi^4$, with coefficient
unity. Being as we are interested in the large $\Lambda$ limit, it will turn out that we can discard the
non-planar term (we could appeal directly to Grosse \& Wulkenhaar's power counting~\cite{G+W-4D} to implement this, but it is easy and instructive to see it directly). First, let us focus on the planar term, in the large $\Lambda$ limit:
\be
	\lim_{\Lambda\rightarrow\infty} 4(n+m)(\gamma_1+\beta_1)
	=
	-\frac{1}{12\pi^2}
	\lim_{\Lambda\rightarrow\infty} 
	\flow
	\left[
	\sum_p
		\hf
		\bigl(
			\Delta(m,p) + \Delta(n,p)
		\bigr)
		+\mbox{non-planar}
	\right]_{m,n}
	,
\ee
where the overall subscript $m,n$ indicates that we are interested in the component linear in either $m$ or $n$. This is given by the next-to-leading term in the discrete Taylor expansion about vanishing $m$ and $n$:
\be
	\lim_{\Lambda\rightarrow\infty}(\gamma_1+\beta_1)
	=
	-\frac{1}{96\pi^2}
	\lim_{\Lambda\rightarrow\infty}
	\flow
	\left[
		\sum_p
		\bigl(
			\Delta(1,p) - \Delta(0,p)
		\bigr)
	\right].
\ee
Writing
\[
	\Delta(1,p) = \frac{K(1/\thetabar) K(p/\thetabar)}{4(3+p)}  
	= \frac{K(p/\thetabar)}{4(3+p)} + \mathcal{O}(1/\thetabar),
	\qquad
	\Delta(0,p) = \frac{ K(p/\thetabar)}{4(2+p)},
\]
it is easy to perform the sum over $p=\{p^1, \ p^2\}$ in the large $\Lambda$ limit. First, replace the sums by integrals with upper limit $\infty$. Then recognize that, to leading order in $1/\thetabar$, we can throw away the cutoff function $K(p/\thetabar)$, so long as the upper limits are reduced to $\thetabar$:
\be
	\lim_{\Lambda\rightarrow\infty}(\gamma_1+\beta_1)
	=
	-\frac{1}{384\pi^2}
	\lim_{\Lambda\rightarrow\infty}
	\flow
	\left[
	\int_0^{\thetabar} dp^1 dp^2
	\left(
		\frac{1}{3+p} - \frac{1}{2+p}
	\right)
	+\mbox{non-planar}
	\right]
\ee
Now, the point about the non-planar term is that, although we expect it to contribute to the \rhs\ (see the discussion about multi-trace terms in \sect{Eff-Ac}), it is sub-leading in $\thetabar$, and hence vanishes in the large $\Lambda$ limit. This can be traced to the fact that there are no integrals to be done in the non-planar case since there are no closed paths in the diagrams which do not hit an external line. Finally, then, we obtain:
\be
	\lim_{\Lambda\rightarrow\infty}(\gamma_1+\beta_1)
	= \frac{1}{192\pi^2}.
\label{eq:gamma_1}
\ee

To obtain a second equation relating $\beta_1$ and $\gamma_1$, we repeat this procedure at the four-point level where we have:
\begin{align}
	\dualv{4}_0 & = \ensuremath{\begin{array}{c}\begin{picture}(0,0)%
\epsfig{file=pstex/S_4_0-nol.pstex}%
\end{picture}%
\setlength{\unitlength}{3947sp}%
\begingroup\makeatletter\ifx\SetFigFont\undefined%
\gdef\SetFigFont#1#2#3#4#5{%
  \reset@font\fontsize{#1}{#2pt}%
  \fontfamily{#3}\fontseries{#4}\fontshape{#5}%
  \selectfont}%
\fi\endgroup%
\begin{picture}(847,847)(4153,-3372)
\put(4533,-3006){\makebox(0,0)[lb]{\smash{{\SetFigFont{12}{14.4}{\rmdefault}{\mddefault}{\updefault}{\color[rgb]{0,0,0}0}%
}}}}
\end{picture}%
 \end{array}},
\\
	\dualv{4}_1 & =  \ensuremath{\begin{array}{c}\begin{picture}(0,0)%
\epsfig{file=pstex/S_4_1-nol.pstex}%
\end{picture}%
\setlength{\unitlength}{3947sp}%
\begingroup\makeatletter\ifx\SetFigFont\undefined%
\gdef\SetFigFont#1#2#3#4#5{%
  \reset@font\fontsize{#1}{#2pt}%
  \fontfamily{#3}\fontseries{#4}\fontshape{#5}%
  \selectfont}%
\fi\endgroup%
\begin{picture}(847,847)(4153,-3372)
\put(4529,-3004){\makebox(0,0)[lb]{\smash{{\SetFigFont{12}{14.4}{\rmdefault}{\mddefault}{\updefault}{\color[rgb]{0,0,0}1}%
}}}}
\end{picture}%
 \end{array}}
	-\frac{1}{12\pi^2}
	\ensuremath{\begin{array}{c}\input{pstex/S_4_0-DD-S_4_0.pstex_t} \end{array}}
	-\frac{1}{5\pi \theta}
	\ensuremath{\begin{array}{c}\begin{picture}(0,0)%
\epsfig{file=pstex/S_6_0-D.pstex}%
\end{picture}%
\setlength{\unitlength}{3947sp}%
\begingroup\makeatletter\ifx\SetFigFont\undefined%
\gdef\SetFigFont#1#2#3#4#5{%
  \reset@font\fontsize{#1}{#2pt}%
  \fontfamily{#3}\fontseries{#4}\fontshape{#5}%
  \selectfont}%
\fi\endgroup%
\begin{picture}(800,884)(4173,-3343)
\put(4534,-3009){\makebox(0,0)[lb]{\smash{{\SetFigFont{12}{14.4}{\rmdefault}{\mddefault}{\updefault}{\color[rgb]{0,0,0}0}%
}}}}
\end{picture}%
 \end{array}}
	+\mbox{non-planar diagrams}.
\label{eq:D^4_1}
\end{align}

The second renormalization condition is that,
ignoring the $1/\hat{\lambda}$ in front of the whole action, the full coefficient of the $1/4! \Tr \phi^4$
term is unity in the $\Lambda\rightarrow\infty$ limit.
This statement is precisely equivalent to saying that, had we not scaled $\sqrt{\hat{\lambda}}$
out of the field, $\hat{\lambda}$ is simply defined to be the coefficient in front of $1/4! \Tr \phi^4$ in the
large-$\Lambda$ limit.
Thus, we can discard the first term on the \rhs\ of~\eq{D^4_1}.
It is worth pointing out that this is where
the renormalization conditions for $\lambda$ and $\hat{\lambda}$ differ: had we rescaled the field
in~\eq{rescale-lambdahat} using $\lambda$ and not $\hat{\lambda}$, then our renormalization condition for the former
would just be that the full coefficient of the $1/4! \Tr \phi^4$
term is unity \emph{at all scales}.

Substituting these diagrammatic expressions into~\eq{D-flow}, and remembering the $1/\hat{\lambda}$ in
\[
	D \sim \frac{1}{\hat{\lambda}} D_0 + D_1 + \mathcal{O}(\hat{\lambda}),
\]
yields:
\be
	-\beta_1 + 4\left(\frac{\gamma_1}{2} + \beta_1\right)
	=
	\flow
	\left[
		\frac{1}{12\pi^2}
		\ensuremath{\begin{array}{c}\input{pstex/S_4_0-DD-S_4_0.pstex_t} \end{array}}
		+\frac{1}{5\pi \theta}
		\ensuremath{\begin{array}{c}\begin{picture}(0,0)%
\epsfig{file=pstex/S_6_0-D.pstex}%
\end{picture}%
\setlength{\unitlength}{3947sp}%
\begingroup\makeatletter\ifx\SetFigFont\undefined%
\gdef\SetFigFont#1#2#3#4#5{%
  \reset@font\fontsize{#1}{#2pt}%
  \fontfamily{#3}\fontseries{#4}\fontshape{#5}%
  \selectfont}%
\fi\endgroup%
\begin{picture}(800,884)(4173,-3343)
\put(4534,-3009){\makebox(0,0)[lb]{\smash{{\SetFigFont{12}{14.4}{\rmdefault}{\mddefault}{\updefault}{\color[rgb]{0,0,0}0}%
}}}}
\end{picture}%
 \end{array}}
		+\mbox{non-planar diagrams}
	\right]_0,
\label{eq:beta_1-pre}
\ee
where the overall subscript 0 instructs us to work to zeroth order in the external indices.
Taking the large $\Lambda$ limit yields:
\be
	\lim_{\Lambda\rightarrow\infty}(2\gamma_1 + 3\beta_1) = 
	\frac{1}{12\pi^2}
	\lim_{\Lambda\rightarrow\infty}
	\flow
	\sum_p \Delta(0,p) \Delta(0,p)	
	=
	\frac{1}{96\pi^2},
\label{eq:beta_1}
\ee
where we have anticipated that only the first diagrams survives the limit (we will justify this in a moment).
Solving~\eqs{gamma_1}{beta_1} we find that:%
\footnote{At first sight, our value for $\gamma_1$ disagrees with
the one found by Grosse \& Wulkenhaar in~\cite{G+W-beta}. This discrepancy arises because  in~\cite{G+W-beta}, $\gamma \equiv \mathcal{N} d\ln \mathcal{Z} /d\mathcal{N}$, where $\mathcal{N} \sim \thetabar$ and $\mathcal{Z} = \sqrt{Z}$. Thus, we should find that $\gamma_{\mathrm{G+W}} = \eta_1/4 = \gamma_1/4$, as indeed we do. Note  that $\eta_1$ and $\gamma_1$ are equal only because $\beta_1$ vanishes [see~\eq{gamma-defn}].}
\be
	\lim_{\Lambda\rightarrow\infty} \gamma_1 = \frac{1}{192\pi^2},
	\qquad
	\lim_{\Lambda\rightarrow\infty} \beta_1 = 0.
\ee

To finish the computation, we must explain why only the first diagram in~\eq{beta_1-pre} survives in the large $\Lambda$-limit. Clearly, the non-planar versions of the first diagram are sub-leading, and so can be discarded. What about the diagram involving the six-point vertex? We cannot throw this term away on the basis that there
is no classical six-point vertex, since the flow equation~\eq{beta-flow} yields:
\be
	-\flow \ensuremath{\begin{array}{c}\begin{picture}(0,0)%
\epsfig{file=pstex/S_6_0.pstex}%
\end{picture}%
\setlength{\unitlength}{3947sp}%
\begingroup\makeatletter\ifx\SetFigFont\undefined%
\gdef\SetFigFont#1#2#3#4#5{%
  \reset@font\fontsize{#1}{#2pt}%
  \fontfamily{#3}\fontseries{#4}\fontshape{#5}%
  \selectfont}%
\fi\endgroup%
\begin{picture}(800,847)(4173,-3375)
\put(4534,-3009){\makebox(0,0)[lb]{\smash{{\SetFigFont{12}{14.4}{\rmdefault}{\mddefault}{\updefault}{\color[rgb]{0,0,0}0}%
}}}}
\end{picture}%
 \end{array}} \propto \frac{\nu_4}{\chi_4} \ensuremath{\begin{array}{c}\input{pstex/S_4-dd-S_4.pstex_t} \end{array}},
\ee
where the dot on the line joining the two vertices on the \rhs\ indicates $\dd$, as opposed to $\Delta$.
(Note, though, that there is no way to generate \emph{classical} four-point functions unless they already exist.)
Now, the effect of this dot, acting on the cutoff functions, is to produce a $1/\thetabar$. Thus we find that
the six-point vertex has a piece which goes like $1/\Lambda^2$ as we could have deduced directly, from dimensional arguments.  However, the six-point vertex also has a piece coming from the integration constant. Given our assumption of self-similarity, this can only go like $\theta \times \mathrm{const}$. Since we are interested in finding renormalized trajectories---for which the action flows
into the Gaussian floating-point as $\Lambda \rightarrow \infty$---we set the constant to zero.

Already, this tells us that the non-planar diagrams involving the six-point vertex vanish, since they go like
\[
	\lim_{\Lambda\rightarrow\infty}
	\flow
	\frac{1}{\thetabar} = 0.
\]
As for the planar diagram, this has the structure
\[
	\lim_{\Lambda\rightarrow\infty}
	\flow
	\left[
		\frac{1}{\thetabar}
		\int_0^{\thetabar}
		dp^1 dp^2
		\frac{1}{2+p^1+p^2}
	\right]
	\sim
	\lim_{\Lambda\rightarrow\infty}
	\frac{\ln \thetabar}{\thetabar}
	=0.
\]

So, at one-loop, the coupling $\hat{\lambda}$ remains marginal (for the self-dual theory); as proven in~\cite{Vanishing}, this remains true to all orders in perturbation theory. Thus, the question of whether or not the Gaussian floating-point supports non-trivial renormalized trajectories is a nonperturbative one, since the exact marginality of 
$\hat{\lambda}$ can in principle be be violated, one way or the other, by exponentially small terms of the form $e^{-1/\hat{\lambda}}$. 
Should the coupling remain exactly marginal then, as discussed extensively around \eq{marginal-criterion}---we expect there to 
exist a line of inequivalent floating-points, connected in one way or another to the Gaussian one. 

Irrespective of this possibility, let us now ask the question: could we make sense of  a trajectory which, from the start, is only specified to pass close to the Gaussian theory, on its journey into the IR (for example if $\hat{\lambda}$ turns out to be nonperturbatively irrelevant)?

The key point is that, with nothing more than this information about the trajectory, we no longer have
any justification in assuming self-similarity (had we found a renormalized trajectory emanating from the Gaussian floating-point, self-similarity would have been justified post hoc).
This would mean that the integration constant for the six-point vertex could have a piece which goes like $1/\Lambda_0^2$. We would now find that the $\beta$-function is corrected by terms which go like $\Lambda^2/ \Lambda_0^2$. 
At first sight, we could remove these terms by sending the bare scale to infinity (before doing likewise with $\Lambda$). However, for this to be a well-defined procedure, it would need to be proven that the remaining perturbative series for the $\beta$-function,  the anomalous dimension and indeed all Wilsonian effective action vertices can be unambiguously resummed.\footnote{In the slightly different context of the multiscale analysis, the proof of resummability in one slice has been given in~\cite{Rivasseau:2007fr}.}
To restate what happens in the commutative case: it is precisely this resummation which cannot be performed; the $\Lambda^2/ \Lambda_0^2$ terms, whose presence can be traced to UV renormalons, must be retained, which violates self-similarity and destroys (nonperturbative) renormalizability.

A way out of this would be, by now needless to say, if a non-trivial floating point exists, which supports a renormalized trajectory that flows towards the Gaussian floating-point.  Note that finding such solutions is not expected to be easy: it would involve solving the flow equation~\eq{Flow}, subject to the floating-point  criterion~\eq{floating}, and then showing that useful trajectories emanate from the putative floating-point. 

\begin{acknowledgments}

We would like to thank Harald Grosse, Daniel Litim, Tim Morris,  Hugh Osborn, Raimar Wulkenhaar and, especially, Francis Dolan and Denjoe O'Connor for useful
discussions. Particular thanks must go to Christian S\"{a}mann, without whom this project would never have got started. O.J.R.\ acknowledges STFC for financial support.  R.G.\ acknowledges the support of the Perimeter Institute.
Research at Perimeter Institute is supported by the Government of Canada through Industry 
Canada and by the Province of Ontario through the Ministry of Research and Innovation.

\end{acknowledgments}

\appendix

\section{The Matrix Base}
\label{app:base}

In this appendix, we review the matrix base in $D=2$  
(following Appendix A of~\cite{G+W-2D} which, in turn, is drawn from~\cite{MatrixBase})
and then construct a basis for the effective action. 
Since the $D=4$ matrix base is built out of two copies
of the $D=2$ base, these results carry over essentially directly to the case of interest.

\subsection{Constructing the Base}
\label{app:basis}

There are two key ingredients to the matrix base. The first of these is the creation and annihilation
operators, defined according to
\be
	a \equiv \frac{1}{\sqrt{2}} (x_1 + i x_2), 
	\qquad 
	\abar \equiv  \frac{1}{\sqrt{2}} (x_1 - ix_2),
\ee
with
\be
	\pder{}{a} = \frac{1}{\sqrt{2}} (\partial_1 - i\partial_2), 
	\qquad 
	\pder{}{\abar} = \frac{1}{\sqrt{2}} (\partial_1 + i\partial_2).
\ee
The creation and annihilations operators satisfy the canonical commutation relation
\be
	[a,\abar]_\star = \theta,\qquad 
\label{eq:can-com}
\ee
where
\be
	[f,g]_\star = f\star g - g\star f, \qquad \{g,f\}_\star = g\star f + f \star g.
\label{eq:star-com}
\ee

Using~\eq{star}, it is a simple matter to check the following relationships, for any $f \in \mathbb{R}^2_\theta$:
\begin{subequations}
\bea
\label{eq:af+deriv}
	(a\star f)(x) = a(x) f(x) + \frac{\theta}{2} \pder{f}{\abar}(x),
	& \qquad &
	(f \star a)(x) = a(x) f(x) - \frac{\theta}{2} \pder{f}{\abar}(x)
\\
\label{eq:abarf+deriv}
	(\abar \star f)(x) = \abar (x) f(x) -  \frac{\theta}{2} \pder{f}{a}(x)	,
	& \qquad &
	(f \star \abar)(x) = a(x) f(x) + \frac{\theta}{2} \pder{f}{a}(x).
\eea
\end{subequations}
From these equations, we deduce the following:
\begin{subequations}
\bea
	a(x) f(x) = \hf \{a, f \}_\star(x), 
	&\qquad&
	\abar (x) f(x) = \hf \{ \abar,f \}_\star(x),
\label{eq:af}
\\
	\pder{f}{a}(x) = - \frac{1}{\theta} [\abar,f]_\star (x) ,
	&\qquad&
	\pder{f}{\abar}(x) = \frac{1}{\theta} [a,f]_\star (x) .
\label{eq:df}
\eea
\end{subequations}
These relationships imply that (with $\alpha = \{1,2\}$)
\be
	\tilde{x}_\alpha f(x) = \hf \{ \tilde{x}_\alpha , f\}_\star(x),
	\qquad
	-\pder{f}{x_\alpha} = \frac{i}{2} [\tilde{x}_\alpha, f]_\star (x),
\ee
as we stated earlier in~\eq{replace}.

Consequently, the the Grosse \& Wulkenhaar action~\eq{G+W}  (which, in $D=4$, contains two copies of the matrix base for $\mathbb{R}^2_\theta$)
can be rewritten such that all pointwise products are replaced by $\star$-products, involving commutators or anticommutators with $\tilde{x}_\alpha$.

Having discussed the creation and annihilation operators, we now introduce the second key ingredient of the matrix base, the Gaussian
\be
	f_0(x) = 2e^{\textstyle -\frac{1}{\theta} (x_1^2  + x_2^2)},
\ee
which is an idempotent:
\be
	(f_0 \star f_0)(x) = f_0(x).
\label{eq:idem}
\ee
When acted upon by the creation and annihilation operators, this function behaves as follows, as
can be checked by using~\eqs{af+deriv}{abarf+deriv}:
\be
	\bar{a}^{\star m} \star f_0 = 2^m \bar{a}^m f_0,
	\qquad
	f_0 \star a^{\star n} = 2^n a^n f_0,
\label{eq:a^nf_0}
\ee
where $a^{\star n} \equiv a \star \cdots \star a$ ($n$ factors) and similarly for $\abar^{\star m}$.

In turn [again using~\eqs{af+deriv}{abarf+deriv}], \eqn{a^nf_0} implies that
\begin{subequations}
\bea
\label{eq:aabar^mf_0}
	a\star \abar^{\star m} \star f_0 
	& = &
	\left\{
	\begin{array}{cl}
		m \theta \abar^{\star (m-1)} \star f_0 & \mbox{for $m \geq 1$}
	\\
		0 & \mbox{for $m = 0$}
	\end{array}
	\right.
\\
\label{eq:f_0abara}
	f_0 \star a^{\star n} \star \abar
	& = &
	\left\{
	\begin{array}{cl}
		n \theta f_0 \star {a}^{\star (n-1)}  & \mbox{for $n \geq 1$}
	\\
		0 & \mbox{for $n = 0$}.
	\end{array}
	\right.
\eea
\end{subequations}

At this point, we are ready to define the basis functions $f_{mn}(x_1,x_2)$, which we have seen already
in~\eq{adapt}:
\bea
\label{eq:f_mn}
	f_{mn} & \equiv &  \frac{1}{\sqrt{n!m!\theta^{n+m}}}
			\abar^{\star m} \star f_0 \star a^{\star n}
\\
\nonumber
			& = &
			\frac{1}{\sqrt{n!m!\theta^{n+m}}}
			\sum_{k=0}^{\mathrm{min}(m,n)}
			(-1)^k 
			\nCr{m}{k}	 \nCr{n}{k} k! \, 2^{m+n-2k} \theta^k \abar^{m-k} a^{n-k} f_0,
\eea
where the second line can be proven by induction, using~\eqs{af+deriv}{abarf+deriv}.

Using~\eq{idem}, \eqs{aabar^mf_0}{f_0abara}, it follows that
\be
	(f_{mn} \star f_{kl})(x) = \delta_{nk} f_{ml}(x).
\ee
It is this multiplication rule which means that the $\star$-product translates into matrix multiplication:
\begin{subequations}
\bea
\label{eq:a-matrix}
	a(x) = \sum_{m,n=0}^\infty a_{mn} f_{mn}(x),
	&\qquad&
	b(x) = \sum_{m,n=0}^\infty b_{mn} f_{mn}(x)
\\
\label{eq:ab-matrix}
	\Rightarrow
	(a \star b)(x) = \sum_{m,n=0}^\infty (ab)_{mn} f_{mn}(x),
	&\qquad &
	(ab)_{mn} = \sum_{k=0}^\infty a_{mk} b_{kn},
\eea
\end{subequations}
where the sequences $\{a_{mn} \}$ must be of rapid decay in order that they describe elements
of $\mathbb{R}^2_\theta$~\cite{MatrixBase}:
\[
	\sum_{m,n}^\infty a_{mn} f_{mn} \in \mathbb{R}^2_\theta
	\qquad
	\mathrm{iff}
	\qquad
	\sum_{m,n}^\infty
	\left(
		(2m+1)^{2k} (2n+1)^{2k} \left| a_{mn} \right|^2
	\right)^{1/2}	< \infty,
	\qquad \forall k.
\]
The normalization of $f_{mn}$ is such that
\be
\label{eq:norm}
	\frac{1}{2\pi \theta} \TwoInt{x} f_{mn}(x) = \delta_{mn}.
\ee

We will conclude this section by providing the matrix base expression for $a$, $\abar$, and various
combinations thereof. We start by recognizing that~\eq{a-matrix} implies
\be
	\TwoInt{x} (a \star f_{nm}) (x)=\sum_{p,q} a_{pq} \TwoInt{x} (f_{pq}\star f_{nm}) (x)
	=2\pi\theta a_{pq}\delta_{qn} \delta_{pm} = a_{mn}.
\ee
(Notice that the ordering of the indices on the $f_{nm}$ is \emph{opposite} to that on the $a_{mn}$.)
Hence the matrix element $a_{mn}$ is 
\be
\label{eq:a_mn}
	a_{mn}=\frac{1}{2\pi\theta} \TwoInt{x} [a\star f_{nm}](x)	
	=\frac{1}{2\pi\theta} \TwoInt{x} \sqrt{(m+1)\theta}f_{nm+1}
	=\sqrt{(m+1)\theta} \; \delta_{m+1n},
\ee
where we have exploited the trace property of the integral and the definition~\eq{f_mn}.
Similarly,
\be
\label{eq:abar_mn}
	\abar_{mn} =\sqrt{m \theta} \delta_{mn+1}.
\ee

From~\eqs{a_mn}{abar_mn}, it follows that
\begin{align}
\label{eq:abara}
	[a \abar]_{mn} & = a_{mp}\bar a_{pn}
			 = \sqrt{(m+1)\theta}\delta_{m+1p}\sqrt{p\theta} \delta_{pn+1}
			= (m+1)\theta \delta_{mn}
\nonumber
\\{}
	[\abar a]_{mn}
	& = \abar_{mp} a_{pn}
		= \sqrt{m \theta} \delta_{m p+1} \sqrt{(p+1)\theta} \delta_{p+1n}
		=  m \theta \delta_{mn}
\end{align}
which, incidentally, gives a direct proof of the fact that $[a,\abar]_{\star}=\theta$. 

Multiplying $a$ (respectively $\abar$) by itself $r$ (respectively $s$) times yields
\bea 
\label{eq:smen}
\nonumber
	[a^{r}]_{mn} 
		& = & a_{mp_1}a_{p_1p_2}\dots a_{p_{r-1}n}
			=\sqrt{(m+1)(p_1+1)\dots(p_{r-1}+1)\theta^{r}}\delta_{m+1p_{1}}
			\delta_{p_1+1p_2}\dots \delta_{p_{r-1}+1 n} 
\\  \nonumber
		& = &\sqrt{\frac{n!}{m!}\theta^r}\delta_{m+r\,n} 
\\ \nonumber
	[\bar a^{s}]_{mn}
		& = &\bar a_{mp_1}\bar a_{p_1p_2}\dots \bar a_{p_{s-1}n} 
		=\sqrt{m p_1 \dots p_{s-1}\theta^{s}} \delta_{mp_{1}+1} 
		\delta_{p_1p_2+1}\dots \delta_{p_{r-1} n+1} 
\\
		& = &\sqrt{\frac{m!}{n!}\theta^s}\delta_{m\,n+s}.
\eea

\subsection{A Basis for the Effective Action}
\label{app:Action-basis}

To illustrate the main points concerning a basis for the effective action, we will continue to work
in $D=2$, and will additionally focus on the two-point part of the action. Writing $\nu_2 \equiv 2\pi\theta$
we have (before any rescalings with $\Lambda$):
\be
\label{eq:S2gen}
	S^{\mathrm{2pt}}= \nu_2
	\sum_{m,n,k,l}
	\hf
	\phi_{mn} 
	A_{mn;kl} \phi_{kl},
\ee
for some arbitrary operator kernel $A_{mn;kl}$ (henceforth, the limits on the various sums are always understood to be from zero to infinity). We will define a procedure to develop this operator on a basis given by traces of products of $a, \bar a$ and $\phi$.

The operator $A_{mn;kl}$ develops on its matrix elements as
\begin{multline}
	A_{mn;kl} = \sum_{r,s \ge 0}\Big{[} 
		A_{mn;n+r\, m+s}\delta_{n+r\,k}\delta_{l\,m+s}+A_{m\,k+r;k\,m+s}\delta_{n\,k+r}\delta_{l\,m+s}
\\
		+ A_{l+s\,n;n+r\,l} \delta_{n+r\,k} \delta_{l+s\,m} 
		+A_{l+s\,k+r;kl}\delta_{l+s\,m}\delta_{n\,k+r} \Big{]}.
\end{multline}
This development holds in all generality. However, in a theory such that $n-m=k-l$, that is theories which conserve the ``angular momentum'' (which is the case for the Grosse \& Wulkenhaar model) it further simplifies. Namely $r=s$ in the first and fourth terms and $r=s=0$ in the second and third terms. 

We write the general development and leave it to the reader to further simplify it. Substituting~\eq{smen} into the above equation and subsequently substituting the whole lot into~\eq{S2gen}, we write
\be
\begin{split}
	S^{\mathrm{2pt}} =\nu_2  \sum_{r,s} \frac{1}{\theta^{(r+s)/2}}
&
	\hf
	\Biggl\{
	\sum_{m,n}
	\left[
		\frac{m!}{(m+s)!}
		\frac{n!}{(n+r)!}
	\right]^{1/2}
	A_{mn;n+r\,m+s}  \phi_{mn} [a^{r}\phi\bar a^s]_{nm}
\\
&
	+
	\sum_{m,k}
	\left[
		\frac{m!}{(m+s)!} \frac{k!}{(k+r)!} 
	\right]^{1/2}
	A_{m\,k+r;k\,m+s}[\phi \bar a^r]_{mk} [\phi\bar a^s]_{km}
\\
&
	+
	\sum_{n,l}
	\left[
		\frac{l!}{(l+s)!} \frac{n!}{(n+r)!}
	\right]^{1/2}
	A_{l+s\,n;n+r\,l} [a^s\phi]_{ln} [a^{r}\phi]_{nl}
\\
&
	+
	\sum_{k,l}
	\left[
		\frac{l!}{(l+s)!} \frac{k!}{(k+r)!}
	\right]^{1/2}
	A_{l+s\,k+r;kl} [a^s\phi \bar a^r]_{lk} \phi_{kl}
	\Biggl\}.
\end{split}
\ee
Taking the example of the first term above we see that, 
at fixed $r$ and $s$, the vertex coefficient function is a function of only $m$ and $n$. Developing it in a Taylor series\footnote{Note that we use here a continuous Taylor development. Obviously one can chose to use a discrete  development, that is to replace the derivatives by finite differences. This only slightly modifies the numerical values of the brings coefficients $B^{\alpha,\beta}_{r,s}$, but does not change the terms in the series.} we write
\bea
\label{eq:two-point-dev}
&&
	 \left[
	 	\frac{1}{\theta^{r+s}}
		\frac{m!}{(m+s)!}
		\frac{n!}{(n+r)!}
	\right]^{1/2}
	A_{mn;n+r\,m+s}
\nonumber
\\
&&
	=\sum_{\alpha,\beta}\frac{1}{\alpha!\beta! \theta^{\alpha+\beta}} 
	\partial_{m}^{\alpha}\partial_{n}^{\beta}
	\Biggl\{
		 \left[
		 	\frac{1}{\theta^{r+s}}
			\frac{m!}{(m+s)!}
			\frac{n!}{(n+r)!}
		\right]^{1/2}
		A_{mn;n+r\,m+s}
	\Biggr\}
	\Biggr\rvert_{m=n=0} 
	(\theta m)^{\alpha} (\theta n)^{\beta} 
\nonumber
\\
&&=\sum_{\alpha,\beta}B^{\alpha,\beta}_{r,s} (\theta m)^{\alpha} (\theta n)^{\beta}
\eea
whereupon, using~\eq{abara}, its contribution to the two point term can be written
\[
	\nu_2
	\sum_{\alpha,\beta, r, s}
	\hf
	B^{\alpha,\beta}_{r,s}
	\,
	\Tr
	\Big{[}
		(\bar a a)^{\alpha}\phi (\bar a a)^{\beta}a^{r}\phi\bar a^s 
	\Big{]}.
\]
Treating the other terms in a similar way, the full two-point vertex can be written as
\begin{multline}
\label{eq:2pt-final}
	S^{\mathrm{2pt}} = \nu_2
	\sum_{\alpha,\beta, r, s}
	\hf
	\Biggl\{
		B^{\alpha,\beta}_{r,s}
		\,
		\Tr
		\Big{[}
			(\bar a a)^{\alpha}\phi (\bar a a)^{\beta}a^{r}\phi\bar a^s 
		\Big{]}
		+
		C^{\alpha,\beta}_{r,s}
		\,
		\Tr
		\Big{[}
			(\abar a)^\alpha \phi \bar a^r (\abar a)^{\beta} \phi\bar a^s 
		\Big{]}
\\
		+
		D^{\alpha,\beta}_{r,s}
		\,
		\Tr
		\Big{[}
			 (\abar a)^{\alpha} a^s\phi  (\abar a)^{\beta} a^{r}\phi
		\Big{]}
		+
		E^{\alpha,\beta}_{r,s}
		\,
		\Tr
		\Big{[}
			 (\abar a)^\alpha a^s\phi \bar a^r  (\abar a)^\beta \phi
		\Big{]}
	\Biggr\},
\end{multline}
where $C^{\alpha,\beta}_{r,s}$, $D^{\alpha,\beta}_{r,s}$, $E^{\alpha,\beta}_{r,s}$ are picked out from the second, third and fourth lines of~\eq{two-point-dev} in direct analogy with $B^{\alpha,\beta}_{r,s}$.
As before, if both $r$ and $s$ are zero, we understand that only one term in~\eq{2pt-final} should be retained, whereas
if $r$ but not $s$ is zero (or vice-versa), we keep two independent terms, discarding their identical copies.

One can readily translate eq. (\ref{eq:2pt-final}) in terms of dimensionless $\tilde X$ operators by substituting
\bea
a=\frac{\theta \Lambda}{\sqrt{2}}(\tilde X_2-i \tilde X_1) \quad 
\bar{a}=\frac{\theta \Lambda}{\sqrt{2}}(\tilde X_2+i \tilde X_1) \; .
\eea 
 Notice that since $\dd_{mn;kl}$ can be thought of as a two-point vertex, our analysis shows that this can be 
written in our basis, and hence translated in to a function of $\tilde{X}$.

Note, though, that we are only interested in theories for which the action can be written without any
loose Lorentz indices. This translates into restrictions on the two point terms we develop on the basis described above.

\section{Operator Norms}
\label{app:norms}

To simplify notation, we use the following shorthands: the index $M_i$ is taken to represent the pair of
indices $(m_i,n_i)$ (each of which, we recall, stands for a further pair of indices) and we define
\[
	\td_{M;N} \equiv \frac{\Delta_{M;N}}{\chibar_4}.
\]

In this section, we will compute
\be
	\lim_{\Lambda \rightarrow \infty} \norm{\mathcal{O}_{J}^{(\varpi)}}
	\equiv \norm{\mathcal{O}_{J}^{(\varpi)}}_\infty,
\ee
where we will take $J$ to be even (it is easy enough to adapt the following analysis to $J$ odd),
and examine those realizations of $(\varpi)$ which encompass all relevant and marginal terms and certain families of irrelevant terms.

To compute the norm, we recall the inner product~\eq{product-2}. This takes the form
of a functional integral with Gaussian weight and so, upon expanding out the $a$ and the $b$
in terms of the fields, we simply sum over all possible contractions of pairs of fields. To simplify the
subsequent analysis, we notice that
\begin{multline}
	\langle \phi_{M_1} \cdots \phi_{M_{2J}} \rangle_\infty 
	\equiv 
	\lim_{\Lambda \rightarrow \infty}
	\frac{
		\ds
		\int \mathcal{D} \phi e^{-\frac{1}{2} \phi \cdot \td^{-1} \cdot \phi} 
		\phi_{M_1} \cdots \phi_{M_{2J}}
	}
		{
		\ds
		\int \mathcal{D} \phi e^{-\frac{1}{2} \phi \cdot \td^{-1} \cdot \phi}
	}
\\
	= 
	\lim_{\Lambda \rightarrow \infty}
	\left.
	\exp \left(
		\hf \classical{}{\td}{}
	\right)
	\phi_{M_1} \cdots \phi_{M_{2J}}
	\right\vert_{\phi=0}
	=
	\lim_{\Lambda \rightarrow \infty}
	\frac{1}{J!}
	\left(
		\hf \classical{}{\td}{}
	\right)^{J}
	\phi_{M_1} \cdots \phi_{M_{2J}}.
\end{multline}

From~\eq{O_j}, we see that we can write
\be
	\mathcal{O}_{J}^{(\varpi)} =
	\sum_{j=0}^{J/2}
	\frac{(-1)^j}{j!}
	\left(
		\hf \classical{}{\td}{}
	\right)^{J}
	u^{(\varpi)}_{M_1\cdots M_J} \phi_{M_1} \cdots \phi_{M_J},
\ee
where we sum over repeated indices.
Putting everything together we have:
\begin{multline}
	\norm{O_J^{(\varpi)}}^2_\infty
	=
	\lim_{\Lambda \rightarrow \infty}
	u^{(\varpi)}_{M_1\cdots M_J}
	u^{(\varpi)}_{K_1\cdots K_J}
	\sum_{j,k=0}^{J/2}
	\frac{(-1)^{j+k}}{(J-j-k)! j!k!}
	\left(\hf \classical{}{\td}{}\right)^{J-j-k}
\\
	\left[
		\left(
		\hf \classical{}{\td}{}
		\right)^{j}
		\phi_{M_1} \cdots \phi_{M_J}
	\right]
	\left[
		\left(
		\hf \classical{}{\td}{}
		\right)^{k}
		\phi_{K_1} \cdots \phi_{K_J}
	\right].
\end{multline}
Consider
the $J-j-k$ pairs of derivatives. For each pair, either both can act on the contents of the first square brackets, or both can act on the contents of the second square brackets, or one can act on the contents of each. Let us denote by $J-2\alpha$ the number of pairs of these derivatives that fall into the latter class. The biggest value $J-2\alpha$ can take is limited by which ever of the square brackets has the least number of fields remaining, after the derivatives within the square brackets have acted.
Now, inside the first square brackets, after the associated derivatives have acted there are
$J-2j$ $\phi$s left. Similarly, inside the second square brackets there are $J-2k$ $\phi$s. Therefore we
see that
\be
	\max(j,k) \leq \alpha \leq J/2.
\ee
Now, after the $J-2\alpha$ derivatives have acted, there are $J-2j-(J-2\alpha) = 2\alpha-2j$ fields
remaining in the first square brackets and $2\alpha -2k$ remaining in the second. Thus, of the
$2\alpha-j-k$ remaining pairs of derivatives, $\alpha-j$ must strike the contents of the first square bracket, leaving $\alpha -k$ to strike the contents of the second:
\begin{multline}
	\norm{O_J^{(\varpi)}}^2_\infty =
	\lim_{\Lambda \rightarrow \infty}
	u^{(\varpi)}_{M_1\cdots K_J}
	u^{(\varpi)}_{K_1\cdots K_J}
	\sum_{j,k=0}^{J/2}
	\frac{(-1)^{j+k}}{(J-j-k)! j!k!}
	\sum_{\alpha = \max(j,k)}^{J/2}
	\nCr{J-j-k}{J-\alpha} \nCr{2\alpha-j-k}{\alpha-j}
\\
	\left[
		\left(
		\hf \classical{}{\td}{}
		\right)^{\alpha}
		\phi_{M_1} \cdots \phi_{M_J}
	\right]
	\left(
		\frac{\stackrel{\leftarrow}{\partial}}{\partial \phi}
		\knl{\td}
		\frac{\stackrel{\rightarrow}{\partial}}{\partial \phi}
	\right)^{J-2\alpha}
	\left[
		\left(
		\hf \classical{}{\td}{}
		\right)^{\alpha}
		\phi_{K_1} \cdots \phi_{K_J}
	\right].
\end{multline}

This expression greatly simplifies. To see how, we start by rewriting the sums according to
\[
	\sum_{j,k=0}^{J/2}\sum_{\alpha = \max(j,k)}^{J/2}
	=
	\sum_{j=0}^{J/2}
	\left(
		\sum_{k=0}^{j-1}
		\sum_{\alpha = j}^{J/2}
		+
		\sum_{k=j}^{J/2}
		\sum_{\alpha = k}^{J/2}
	\right),
\]
which yields:
\begin{multline}
	\norm{O_J^{(\varpi)}}^2_\infty =
	\lim_{\Lambda \rightarrow \infty}
	u^{(\varpi)}_{M_1\cdots M_J}
	u^{(\varpi)}_{K_1\cdots K_J}
	\sum_{j=0}^{J/2}
	\left(
		\sum_{k=0}^{j-1}
		\sum_{\alpha = j}^{J/2}
		+
		\sum_{k=j}^{J/2}
		\sum_{\alpha = k}^{J/2}
	\right)
	\frac{(-1)^{j+k}}{j!k! (J-2\alpha)! (\alpha-j)!  (\alpha-k)!}
\\
	\left[
		\left(
		\hf \classical{}{\td}{}
		\right)^{\alpha}
		\phi_{M_1} \cdots \phi_{M_J}
	\right]
	\left(
		\frac{\stackrel{\leftarrow}{\partial}}{\partial \phi}
		\knl{\td}
		\frac{\stackrel{\rightarrow}{\partial}}{\partial \phi}
	\right)^{J-2\alpha}
	\left[
		\left(
		\hf \classical{}{\td}{}
		\right)^{\alpha}
		\phi_{K_1} \cdots \phi_{K_J}
	\right]
\end{multline}

Recognizing that
\[
	\sum_{k=j}^{J/2} \sum_{\alpha=k}^{J/2} = \sum_{\alpha=j}^{J/2} \sum_{k=j}^{\alpha}
\]
we have:
\begin{multline}
	\norm{O_J^{(\varpi)}}^2_\infty =
	\lim_{\Lambda \rightarrow \infty}
	u^{(\varpi)}_{M_1\cdots M_J}
	u^{(\varpi)}_{K_1\cdots K_J}
	\sum_{j=0}^{J/2}
	\sum_{\alpha=j}^{J/2}
	\frac{(-1)^j}{j! (\alpha-j)! (J-2\alpha)! \alpha!}
	\sum_{k=0}^\alpha
	(-1)^k \nCr{\alpha}{k}
\\
	\left[
		\left(
		\hf \classical{}{\td}{}
		\right)^{\alpha}
		\phi_{M_1} \cdots \phi_{M_J}
	\right]
	\left(
		\frac{\stackrel{\leftarrow}{\partial}}{\partial \phi}
		\knl{\td}
		\frac{\stackrel{\rightarrow}{\partial}}{\partial \phi}
	\right)^{J-2\alpha}
	\left[
		\left(
		\hf \classical{}{\td}{}
		\right)^{\alpha}
		\phi_{K_1} \cdots \phi_{K_J}
	\right].
\end{multline}
Looking at the sum over $k$, it is apparent that this expression vanishes unless $\alpha = 0$. In turn, 
this implies that we must take $j=0$. Finally, then, we see that the expression for the norm of the $\mathcal{O}$s takes a very simple form:
\begin{align}
\nonumber
	\norm{O_J^{(\varpi)}}^2_\infty
	&=
	\lim_{\Lambda \rightarrow \infty}
	u^{(\varpi)}_{M_1\cdots M_J}
	u^{(\varpi)}_{K_1\cdots K_J}
	\left[
	\phi_{M_1} \cdots \phi_{M_J}
	\frac{1}{J!}
	\left(
		\frac{\stackrel{\leftarrow}{\partial}}{\partial \phi}
		\knl{\td}
		\frac{\stackrel{\rightarrow}{\partial}}{\partial \phi}
	\right)^{J}
	\phi_{K_1} \cdots \phi_{K_J}
	\right].
\\
\nonumber
	& =
	\lim_{\Lambda \rightarrow \infty}
	u^{(\varpi)}_{M_1\cdots M_J}
	u^{(\varpi)}_{K_1\cdots K_J}
	\left[
		\Delta_{M_1; K_1} \cdots \Delta_{M_J; K_J} + \mathrm{permutations}
	\right]
\\
	& = 
	\lim_{\Lambda \rightarrow \infty}
	u^{(\varpi)}_{m_1n_1\cdots m_Jn_J}
	u^{(\varpi)}_{k_1l_1\cdots k_J l_J}
	\sum_{j_1 \neq j_2 \neq \cdots \neq j_J}
	\prod_{i=1}^J
	\frac{c(m_i,n_i;\thetabar)}{4\chibar_4(2+m_i +n_i)} \delta_{m_i l_{j_i}}\delta_{n_i k_{j_i}}
\end{align}

\subsection{Operators built from Kronecker-$\delta$s}

As a first exercise, we will consider the case where $u^{(\varpi)}$ is just a string of $\delta$-functions:
\be
	u^{(\delta)}_{m_1n_1\cdots m_Jn_J} = \prod_{i=1}^J \delta_{n_i m_{i+1}},
\ee
where we identify the index $i+J$ with $i$. The superscript $(\delta)$ denotes the restriction to terms possessing built only out of strings of $\delta$-functions. This leads to:
\begin{align}
\nonumber
	\norm{O_J^{(\delta)}}^2_\infty
	&=
	\lim_{\Lambda \rightarrow \infty}
	\left(
		\prod_{i=1}^{J}
		\delta_{n_i m_{i+1}}
	\right)
	\left(
		\prod_{i=1}^{J}
		\delta_{l_{i} k_{i+1}}
	\right)
	\left[
		\sum_{j_1 \neq j_2 \neq \cdots \neq j_J}
		\prod_{i=1}^J
		\frac{c(m_{i},n_{i};\thetabar)}{4\chibar_4(2+m_{i} +n_{i})} \delta_{m_{i} l_{j_{i}}}
		\delta_{n_{i} k_{j_{i}}}
	\right]
\\
\nonumber
	& =
	\lim_{\Lambda \rightarrow \infty}
	\left[
		\sum_{j_1 \neq j_2 \neq \cdots \neq j_J}
		\prod_{i=1}^J
		\frac{c(m_{i},m_{i+1};\thetabar)}{4\chibar_4(2+m_{i} +m_{i+1})} 
		\delta_{m_{i} k_{j_{i}+1}}\delta_{m_{i+1} k_{j_{i}}}
	\right]
\\
	& =
	\lim_{\Lambda \rightarrow \infty}
	\left[
		\sum_{j_1 \neq j_2 \neq \cdots \neq j_J}
		\prod_{i=1}^J
		\frac{c(k_{j_{i}+1},k_{j_{i}};\thetabar)}{4\chibar_4(2+k_{j_{i}+1} +k_{j_{i}})} 
		\delta_{k_{j_{i+1}+1} k_{j_{i}}}
	\right]
\label{eq:lim-prelim}
\end{align}
where, in the last expression, we must remember that we are summing over $k_1, \ldots, k_J$.

We now wish to compute the leading behaviour in $\thetabar$ in the $\Lambda \rightarrow \infty$ limit.
To do this we will, at the appropriate juncture, replace each index sum with an integral, 
the upper limits of which are $\thetabar$ (the difference between the sum and the integral is, in this limit, subleading).  However, it may be that some of these nascent integrals are killed (before they actually come into being) by the Kronecker-$\delta$s. Since we are interested in the leading behaviour in $\thetabar$, we are thus interested in the
case where the minimum number of integrals is killed. In fact, we need not kill any; this can be achieved by taking
\[
	j_2+1 = j_1,\qquad j_3+1 = j_2, \qquad \ldots \qquad j_{J+1} + 1 = j_J,
\]
since then the product of Kronecker-$\delta$s in~\eq{lim-prelim} becomes
\[
	\prod_{i=1}^J \delta_{k_i k_i}.
\]
Thus we have:
\begin{align}
\nonumber
	\norm{O_J^{(\delta)}}^2_\infty
	& =
	\lim_{\Lambda \rightarrow \infty}
	\sum_{k_1,\ldots, k_J}
	\sum_{j=1}^J
	\prod_{i=0}^{J-1}
	\frac{c(k_{j-i},k_{j-i-1};\thetabar)}{4\chibar_4(2+k_{j-i} + k_{j-i-1})}
\\
	& =
	\lim_{\Lambda \rightarrow \infty}
	\frac{J}{(4\pi)^{2J}}
	\frac{1}{\bar{\theta}^J}
	\int_0^{\thetabar} dk_1 \cdots dk_J
	\prod_{i=1}^{J}
	\frac{1}{2+k_{i} + k_{i-1}},
\label{eq:norm-delta}
\end{align}
where, as usual, identify $k_0$ with $k_J$.
Recalling that the $k_i$ represent two indices, we see by power counting that we expect
\be
	\norm{O_J^{(\delta)}}^2_\infty = 
	\lim_{\Lambda \rightarrow \infty} 
	\left(A_J \ln \thetabar + B_J\right),
\label{eq:delta-lim}
\ee
We will now show that $A_J$ vanishes.

To do this, we first combine the denominators in~\eq{norm-delta} using Feynman parameters, $x_i$:
\begin{multline}
	\norm{O_J^{(\delta)}}^2_\infty
\\
	\sim
	\lim_{\Lambda \rightarrow \infty}
	\frac{1}{\bar{\theta}^J}
	\int_0^{\thetabar} dk_1\cdots dk_J
	\int_0^1 dx_1\cdots dx_J \delta(1-x_1-\cdots-x_J) 
	\frac{(J-1)!}{(2+k_1b_1 + \cdots + k_J b_J)^J}
\label{eq:starting}
\end{multline}
where, for brevity, we have dropped the overall constant $J/(4\pi)^{2J}$ and we define
\be
	b_i \equiv x_i + x_{i+1}.
\ee
We now perform the $J$ integrals over the $k_i^1$s to yield
\begin{align}
\nonumber
	\norm{O_J^{(\delta)}}^2_\infty
	& \sim
	\lim_{\Lambda \rightarrow \infty}
	\frac{(-1)^{J-1}}{\bar{\theta}^J}
	\int_0^{\thetabar} dk_1^2\cdots dk_J^2
	\int_0^1 dx_1\cdots dx_J \delta(1-x_1-\cdots-x_J) 
	\frac{1}{b_1 b_2 \cdots b_J}
\\
	& =
	\sum_{i_1=0}^1\cdots \sum_{i_J=0}^1
	(-1)^{(i_1+1)+\cdots+(i_J+1)}
	\ln
	\left[
		2 + b_1 k_1^2 + \cdots b_J k_J^2 + (i_1b_1 +\cdots i_jb_J)\thetabar
	\right].
\end{align}

To perform the integrals over the $k_i^2$s, we need the result that
\be
	\int_0^{\thetabar} dy
	(a + by)^n \left[\ln(a+by) - \zeta_n(1) \right]
	=
	\frac{1}{b}
	\left[
		\frac{y^{n+1}}{n+1} \bigl(\ln y - \zeta_{n+1}(1)\bigr)
	\right]^{a+b\thetabar}_0,
\ee
where
\be
	\zeta_n(1) \equiv \sum_{i=1}^n \frac{1}{n},
\ee
and we identify $\zeta_0(1) \equiv 0$.

As a warm up, let us define
\be
	A_j \equiv 2 + k_j^2 b_j + \cdots + k_J^2 b_J  + (i_1b_1 +\cdots i_jb_J)\thetabar
\ee
and consider
\begin{align}
\nonumber
	\int_0^{\thetabar} dk_1^2 \ln (A_2 + b_1k_1^2)
	& =
	\frac{1}{b_1}
	\biggl[
		(A_2 + b_1 \thetabar) \left[\ln (A_2 + b_1 \thetabar) -1\right] - A_2 (\ln A_2 -1)
	\biggr]
\\
	& =
	\frac{1}{b_1}
	\sum_{j_1=0}^1 (-1)^{(j_1+1)} (A_2 +j_1 b_1 \thetabar) 
	\left[
		\ln (A_2 +j_1 b_1 \thetabar) - \zeta_1(1)
	\right].
\end{align}
It is now a simple matter to see that
\begin{multline}
	\norm{O_J^{(\delta)}}^2_\infty 
	\sim
	\lim_{\Lambda \rightarrow \infty}
	\frac{(-1)^{J+1}}{\bar{\theta}^J}
	\int_0^1 dx_1\cdots dx_J \delta(1-x_1-\cdots-x_J) 
	\frac{1}{(b_1 b_2 \cdots b_J)^2}
\\
	\sum_{i_1=0}^1 \cdots \sum_{i_J=0}^1\sum_{j_1=0}^1 \cdots \sum_{j_J=0}^1
	(-1)^{i_1 + \cdots + i_J + j_1 +\cdots +j_J}
	\left[
		2 + \bigl( (i_1+j_1) b_1 + \cdots + (i_J+j_J) b_J \bigr) \thetabar
	\right]^J
\\
	\left[
		\ln 
		\Biggl(
			2+ 
			\bigl(
				(i_1+j_1)b_1 + \cdots + (i_J+j_J)b_J
			\bigr)
			\thetabar
		\Biggr)
		-\zeta_J(1)
	\right].
\end{multline}
As anticipated in~\eq{delta-lim}, the leading behaviour in the $\Lambda \rightarrow \infty$ limit is
\[
	\norm{O_J^{(\delta)}}^2_\infty
	= \lim_{\Lambda \rightarrow \infty} \left(A_J \ln \thetabar + B_J\right)
\]
where, up to an unimportant constant,
\begin{align}
\nonumber
	A_J
	&
	\sim \int_0^1 dx_1\cdots dx_J \delta(1-x_1-\cdots-x_J) 
	\frac{1}{(b_1 b_2 \cdots b_J)^2}
\\
	&
	\qquad
	\sum_{i_1=0}^1 \cdots \sum_{i_J=0}^1\sum_{j_1=0}^1 \cdots \sum_{j_J=0}^1
	(-1)^{i_1 + \cdots + i_J + j_1 +\cdots +j_J}
	\left[
		(i_1+j_1) b_1 + \cdots + (i_J+j_J) b_J
	\right]^J.
\end{align}
Let us focus on the final line, which we can write as:
\begin{multline}
	\sum_{i_1=0}^1 \cdots \sum_{i_J=0}^1\sum_{j_1=0}^1 \cdots \sum_{j_J=0}^1
	(-1)^{i_1 + \cdots + i_J + j_1 +\cdots +j_J}
\\ 
	\sum_{\alpha_1 + \cdots + \alpha_J = J}
	\nCr{J}{\alpha_1} \nCr{J-\alpha_1}{\alpha_2} \cdots \nCr{J-\alpha_1-\cdots -\alpha_{J-1}}{\alpha_J}
	\left[
		(i_1 + j_1) b_1
	\right]^{\alpha_1}
	\cdots
	\left[
		(i_J + j_J) b_J
	\right]^{\alpha_J}.
\end{multline}
Now, if $\alpha_k = 0$, for any $k$, then the expression vanishes, since
\[
	\sum_{i_k=0}^1 (-1)^{i_k} = 0.
\]
Therefore, we must have that $\alpha_k = 1, \  \forall k$. But, since
\be
	\sum_{i_k=0}^1 \sum_{j_k=0}^1 (-1)^{i_k + j_k} (i_k + j_k) = 0,
\label{eq:vanish}
\ee
this contribution vanishes too. Therefore,
\be
	A_J = 0.
\ee
As for $B_J$, the contributions coming from the $\zeta_J(1)$ vanish, for exactly the same reason that $A_J$ vanishes. Reinserting the overall constant gives:
\begin{multline}
	B_J =
	\frac{(-1)^{J+1}J}{(4\pi)^{2J}}
	\int_0^1 dx_1\cdots dx_J \delta(1-x_1-\cdots-x_J) 
	\frac{1}{(b_1 b_2 \cdots b_J)^2}
	\sum_{i_1=0}^1 \cdots \sum_{i_J=0}^1\sum_{j_1=0}^1 \cdots \sum_{j_J=0}^1
\\
	(-1)^{i_1 + \cdots + i_J + j_1 +\cdots +j_J}
		\bigl[ (i_1+j_1) b_1 + \cdots + (i_J+j_J) b_J \bigr]^J
		\ln 
			\bigl[
				(i_1+j_1)b_1 + \cdots + (i_J+j_J)b_J
			\bigr].
\end{multline}
The first $B_J$ is:
\be
	B_2 = \frac{8}{(4\pi)^4}
	(
	9 \ln 3 
	-
	14 \ln 2
	).
\ee

\subsection{Adding $\tilde{X}$s}

Having shown that operators constructed out of just Kronecker-$\delta$s have constant norm,
we now move on to consider more complicated cases. All the operators in our theory can be built
by sandwiching $\tilde{X}_\mu$s between the various $\phi$s. Now, since all Lorentz indices must
be contracted, we are led to consider%
\footnote{
If we were to contract  $(\tilde{X}_\mu)_{mn}$ and $(\tilde{X}_\nu)_{kl}$ via $\thetabar^{\mu \nu}$ (or its inverse), the result will be antisymmetric under interchange of $(m,n)$ with $(k,l)$. Since interactions
are invariant under permutations of the fields, such contributions die and so need not be considered
any further.}
\begin{multline}
	\bigl( \tilde{X}_\mu \bigr)_{mn}  \bigl( \tilde{X}^\mu \bigr)_{kl} 
	=
	\frac{4}{\thetabar}
	\Bigl[
	\left(
		\sqrt{m^1 l^1} \delta_{m^1 n^1+1} \delta_{k^1+1 l^1} 
		+\sqrt{n^1 k^1} \delta_{m^1+1 n^1} \delta_{k^1 l^1+1}
	\right)
	\delta_{m^2l^2} \delta_{n^2k^2}
	\\
	+
	\left(
		\sqrt{m^2 l^2} \delta_{m^2 n^2+1} \delta_{k^2+1 l^2} 
		+\sqrt{n^2 k^2} \delta_{m^2+1 n^2} \delta_{k^2 l^2+1}
	\right)
	\delta_{m^1l^1} \delta_{n^1k^1}
	\Bigr],
\label{eq:XX}
\end{multline}
as the basic ingredient for what follows.

Now, to begin with, we will suppose that all instances of $\tilde{X}$ are such that, with the above
parametrization, $n=k$ (we will deal with the more general case, later). In this case we have:
\be
	\bigl( \tilde{X}_\mu \tilde{X}^{\mu} \bigr)_{ml} = \frac{4}{\thetabar} (1 + m) \delta_{ml}.
\ee
Considering, as an example, an operator which has a piece like
\[
	\phi_{m_1 n_1} \bigl( \tilde{X}_\mu \tilde{X}^{\mu} \bigr)_{n_1m_2}
	\phi_{m_2 n_2} \cdots,
\]
it is clear that the the operators we are considering are specified by
\be
	u^{(\kappa_1,\ldots,\kappa_J)}_{m_1n_1\cdots m_Jn_J} = 
	\sum_{\{\alpha_1,\ldots,\alpha_J\} = \mathrm{cyclic}\{\kappa_1,\ldots,\kappa_J\}}
	\Biggl[ 
		\prod_{i=1}^J \delta_{n_i m_{i+1}} \Biggl(\frac{1+m_{i+1}}{\thetabar} \Biggr)^{\alpha_i} 
		4^{\alpha_i}
	\Biggr],
\ee
where the $\kappa_i$ are non-negative integers, and the ordered set of $\alpha_i$
take values specified by the independent cyclic permutations of the ordered $\kappa_i$.
The purpose of the peculiar looking sum to symmetrize over contributions which are
identical as a consequence of the trace structure of the interaction. For example,
\[
	 \phi_{m_1n_1} \bigl(\tilde{X}_\mu  \tilde{X}^\mu\bigr)_{n_1 m_2} \phi_{m_2 m_1}	
	=
	 \phi_{m_1n_1} \phi_{n_1 m_2} \bigl(\tilde{X}_\mu  \tilde{X}^\mu\bigr)_{m_2 n_1}. 
\]
As a shorthand for the full operators, we will use $\mathcal{O}^{(\kappa)}_J$.

The computation of the norm in the large $\Lambda$-limit is similar to before, but with an obvious difference: when we convert the sums to integrals, there will be extra powers of $k_i$ in the numerator.
Due to the $1/\thetabar$ which accompanies each $(1+m_{i+1})$, simple power counting indicates
that the only terms which survive in the large-$\Lambda$ limit are those for which the constant piece
(\ie\ unity) is dropped. In this case, after converting the sums to integrals, every power of $k_i$ in the numerator comes with precisely one factor of $1/\thetabar$; as in the previous case, the norm is a constant, up to possible logarithmic pieces. We now show that, once again, the logarithmic pieces vanish.

The type of integrals we are now interested in look like
\be
	F_J^{(\alpha_1,\ldots,\alpha_J)} =
	\frac{1}{\thetabar^{J+\alpha_1 +\cdots+ \alpha_J}}
	\int_0^{\thetabar} dk_1\cdots dk_J 
	\frac{k_1^{\alpha_1} \cdots k_J^{\alpha_J}}{(2+k_1b_1 + \cdots + k_J b_J)^J}.
\ee
(Of course, we must at some point integrate over the Feynman parameters, but this is not
necessary to show that the contributions to the norms logarithmic in $\thetabar$ vanish.)

To compute this integral, let us define
\be
	G_{I-J} \equiv
	\int^0 dy_1 \cdots dy_I
	(2+k_1b_1 + \cdots + k_J b_J + y_1 + \cdots+  y_I)^{-J}.
\ee
Now we can write
\be
	F_J^{(\alpha_1,\ldots,\alpha_J)} 
	=
	\frac{1}{\thetabar^{J+\alpha_1 +\cdots+ \alpha_J}}
	\int_0^{\thetabar} dk_1\cdots dk_J 
	\Biggl(
		\pder{}{b_1}
	\Biggr)^{\alpha_1}
	\cdots
	\Biggl(
		\pder{}{b_J}
	\Biggr)^{\alpha_J}
	G_{\alpha_1+\cdots + \alpha_J-J}
\ee
Interchanging the order of differentiation and integration we perform the integrals over the $k_i$ (for which we can read the result off from the last section), followed by the integrals over the $y_i$ to yield:
\begin{multline}
	F_J^{(\alpha_1,\ldots,\alpha_J)}  \sim
	\frac{1}{\thetabar^{J+\alpha_1 +\cdots+ \alpha_J}}
	\sum_{i_1=0}^1 \cdots \sum_{i_J=0}^1\sum_{j_1=0}^1 \cdots \sum_{j_J=0}^1
	(-1)^{i_1 + \cdots + i_J + j_1 +\cdots +j_J}
	\Biggl(
		\pder{}{b_1}
	\Biggr)^{\alpha_1}
	\cdots
	\Biggl(
		\pder{}{b_J}
	\Biggr)^{\alpha_J}
\\
	\frac{1}{(b_1 b_2 \cdots b_J)^2}
	\left[
		2 + \bigl( (i_1+j_1) b_1 + \cdots + (i_J+j_J) b_J \bigr) \thetabar
	\right]^{J+\alpha_1 + \cdots + \alpha_J}
\\
	\left[
		\ln 
		\Biggl(
			2+ 
			\bigl(
				(i_1+j_1)b_1 + \cdots + (i_J+j_J)b_J
			\bigr)
			\thetabar
		\Biggr)
		-\zeta_{J+\alpha_1+\cdots +\alpha_J}(1)
	\right].
\end{multline}
The leading behaviour in the $\Lambda \rightarrow \infty$ limit is:
\begin{multline}
	\lim_{\Lambda \rightarrow \infty}
	F_J^{(\alpha_1,\ldots,\alpha_J)}  
	\sim
	(J+\alpha_1+\cdots+ \alpha_J)! \ln \thetabar
	\sum_{i_1=0}^1 \cdots \sum_{i_J=0}^1\sum_{j_1=0}^1 \cdots \sum_{j_J=0}^1
	(-1)^{i_1 + \cdots + i_J + j_1 +\cdots +j_J}
\\	
	\sum_{\beta_1 = 0}^{\alpha_1} \cdots \sum_{\beta_J = 0}^{\alpha_J}
	\nCr{\alpha_1}{\beta_1} \cdots \nCr{\alpha_J}{\beta_J} (-1)^{\beta_1 + \cdots + \beta_J}
	(\beta_1+1)!\cdots (\beta_J+1)!
\\
	\sum_{\gamma_1=0}^{J+\beta_1+\cdots+\beta_J} \cdots
	\sum_{\gamma_J=0}^{J+\beta_1+\cdots+\beta_J}
	\delta_{J+\beta_1+\cdots+\beta_J-\gamma_1-\cdots-\gamma_J}
\\
	\frac{1}{b_1^{2+\beta_1-\gamma_1} \cdots b_J^{2+\beta_J-\gamma_J}}
	\frac{1}{\gamma_1! \cdots \gamma_J!} 
	(i_1+j_1)^{\alpha_1-\beta_1+\gamma_1}
	\cdots
	(i_J+j_J)^{\alpha_J-\beta_J+\gamma_J}.
\end{multline}
Defining
\be
	\epsilon_i \equiv \gamma_i - \beta_i,
\ee
we can write:
\begin{multline}
	\lim_{\Lambda \rightarrow \infty}
	F_J^{(\alpha_1,\ldots,\alpha_J)}  
	\sim
	(J+\alpha_1+\cdots+ \alpha_J)! \ln \thetabar
	\sum_{i_1=0}^1 \cdots \sum_{i_J=0}^1\sum_{j_1=0}^1 \cdots \sum_{j_J=0}^1
	(-1)^{i_1 + \cdots + i_J + j_1 +\cdots +j_J}
\\	
	\sum_{\beta_1 = 0}^{\alpha_1} \cdots \sum_{\beta_J = 0}^{\alpha_J}
	\sum_{\epsilon_1=-\beta_1}^{J+\beta_2+\cdots+\beta_J}
	\cdots \sum_{\epsilon_J=-\beta_J}^{J+\beta_1+\cdots+\beta_{J-1}}
	\delta_{J-\epsilon_1-\cdots-\epsilon_J}
	\frac{1}{b_1^{2-\epsilon_1} \cdots b_J^{2+\epsilon_J}}
	(i_1+j_1)^{\alpha_1+\epsilon_1}
	\cdots
	(i_J+j_J)^{\alpha_J+\epsilon_J}
\\	
	\nCr{\alpha_1}{\beta_1} \cdots \nCr{\alpha_J}{\beta_J} (-1)^{\beta_1 + \cdots + \beta_J}
	\frac{(\beta_1+1)!}{(\beta_1+\epsilon_1)!}
	\cdots 
	\frac{(\beta_J+1)!}{(\beta_J+\epsilon_J)!}.
\end{multline}
As we might guess, it turns out that this expression vanishes. To see how, let us start by supposing that
one of the $\epsilon$s is unity, say $\epsilon_1$. But now the sum over $\beta_1$ vanishes, unless $\alpha_1=0$. However, if $\alpha_1=0$, then the sum over $i_1$ (and $j_1$) causes the expression as a whole to vanish.

Now, given that the $\epsilon_i$ sum to $J$, it is clear that if none of the $\epsilon_i$ can equal unity, then at least one of the $\epsilon$s must be non-positive. Taking this to be $\epsilon_1 = -\delta_1$, we are led to consider
\be
\begin{split}
	&
	\sum_{i_1=0}^1 \sum_{j_1=0}^1 (-1)^{i_1+j_1} 
	\sum_{\beta_1=0}^{\alpha_1}
	\sum_{\delta_1= 0}^{\beta_1}
	(i_1+j_1)^{\alpha_1- \delta_1}
	 \nCr{\alpha_1}{\beta_1} (-1)^{\beta_1}
	\frac{(\beta_1+1)!}{(\beta_1- \delta_1)!}
\\
	=
	&
	\sum_{i_1=0}^1 \sum_{j_1=0}^1 (-1)^{i_1+j_1} 
	\sum_{\delta_1= 0}^{\alpha_1}
	\sum_{\beta_1=\delta_1}^{\alpha_1}
	(i_1+j_1)^{\alpha_1- \delta_1}
	 \nCr{\alpha_1}{\beta_1} (-1)^{\beta_1}
	\frac{(\beta_1+1)!}{(\beta_1- \delta_1)!}.
\\
	=
	&
	\sum_{i_1=0}^1 \sum_{j_1=0}^1 (-1)^{i_1+j_1} 
	\sum_{\delta_1= 0}^{\alpha_1}
	\frac{\alpha_1!}{(\alpha_1-\delta_1)!}
	(i_1+j_1)^{\alpha_1- \delta_1}
	\sum_{\beta_1=0}^{\alpha_1-\delta_1}
	\nCr{\alpha_1-\delta_1}{\beta_1} (\beta_1+\delta_1+1)
\end{split}
\ee

However, using the
result that
\be
	\sum_{p=0}^n \nCr{n}{p} p^q =
	\left\{
		\begin{array}{rl}
			0 & q<n
		\\
			\neq 0 & q\geq n
		\end{array}
	\right.,
\ee
it is apparent that we must take $\alpha_1 -\delta_1= 0$ or $1$. In the first case, the sums over $i_1$ and $j_1$ separately cause the expression to vanish; in the latter case, the combined sum over $i$ and $j$ causes the expression to vanish.

In conclusion, then, it follows that
\be
	\norm{\mathcal{O}^{(\kappa)}_{J}}_\infty = \mathrm{const}.
\ee

Next, we will look at the norm of an operator possessing a pair of $\tilde{X}$s which are separated,
\viz\
\be
	\phi_{m_1n_1} \bigl(\tilde{X}_\mu\bigr)_{n_1m_2} \phi_{m_2 m_3} \cdots
	\phi_{m_In_I}  \bigl(\tilde{X}^\mu\bigr)_{n_I m_{I+1}} \phi_{m_{I+1} m_{I+2}}
	\cdots \phi_{m_J m_1}.
\label{eq:XsplitX}
\ee
When computing the norm, we will start by looking at the contributions from
$\bigl(\tilde{X}_\mu\bigr)_{n_1m_2}  \bigl(\tilde{X}^\mu\bigr)_{n_I m_{I+1}}$ that have non-trivial
structure in the indices $n_1^1,\ m_2^2,\ n_I^1$ and $m_{I+1}^1$. Thus, looking at~\eq{XX}, we will focus on the pair of terms in the first
round brackets. 
Defining
\be
	r_i = 
	\left\{
		\begin{array}{rl}
			1 & i=1
		\\
			-1 & i=I
		\\
			0 & \mathrm{otherwise},
		\end{array}
	\right.
\label{eq:r_i}
\ee
we have the following contribution to the norm coming from the indices with a superscript 1:
\begin{multline}
	\frac{1}{\thetabar}
	\left[
		\left(\prod_{i=1}^J \delta_{n_i^1 m_{i+1}^1 + r_i}\right)
		(n_1^1 m_{I+1}^1)^{1/2}
		+
		\left(\prod_{i=1}^J \delta_{n_i^1 + r_i m_{i+1}^1 }\right)
		(n_I^1 m_2^1)^{1/2}
	\right]
\\
	\left[
		\left(\prod_{i=1}^J \delta_{l_i^1 k_{i+1}^1 + r_i}\right)
		(l_1^1 k_{I+1}^1)^{1/2}
		+
		\left(\prod_{i=1}^J \delta_{l_i^1 + r_i k_{i+1}^1 }\right)
		(l_I^1 k_2^1)^{1/2}
	\right]
	\sum_{j_1\neq \cdots \neq j_J}
	\left(
		\prod_{i=1}^J \delta_{m_i^1 l_{j_i}^1} \delta_{n_i^1 k_{j_i}^1}
	\right).
\label{eq:awkward}
\end{multline}

Let us focus on the term formed by taking the first contribution from each of the pairs of square brackets.
Performing the sums over the $n_i,\ l_i$ and $m_i$ yields:
\be
	\frac{1}{\thetabar}
	\sum_{j_1\neq \cdots \neq j_J}
	\delta_{k^1_{j_{i+1}+1} + r_{j_{i+1}} + r_i \; k_{j_i}^1}
	\left[
		k_{j_1}^1 (k^1_2 + r_1) (k^1_{j_I} - r_I) k^1_{I+1}
	\right]^{1/2}.
\label{eq:sqrt}
\ee
Now, the leading term in the large-$\Lambda$ limit occurs when the Kronecker-$\delta$s do not kill
any of the sums of the $k_i$. This requires that
\be
	j_{i+1} + 1 = j_i, \qquad r_{j_{i+1}} = -r_i.
\ee
Using~\eq{r_i}, we therefore deduce that
\begin{align}
	r_{j_2} & = -r_1 = 1, \qquad\Rightarrow j_2 = I, \qquad \Rightarrow j_1 = I+1;
\\
	r_{j_{I+1}} & = -r_I = 1, \qquad \Rightarrow j_{I+1} = 1, \qquad \Rightarrow j_{I} = 2,
\end{align}
from which it is apparent that the argument of the square root in~\eq{sqrt} is a perfect square.

Consequently, we are left with terms like those we found in our analysis of the operators $\mathcal{O}^{(\kappa)}_J$: the sums over the $k_i^1$, which become integrals, have leading terms in the numerator like $k_i^1 k_j^1 /\thetabar$. We know that such contributions give rise to a constant norm. It is easy to check that the same thing happens for all contributions coming from~\eq{awkward}. Thus, the operator
corresponding to~\eq{XsplitX} has constant norm.

\bibliography{ERG,NC}

\end{document}